\documentclass[sigconf]{acmart} 
\AtBeginDocument{%
  }

\copyrightyear{2026}
\acmYear{2026}
\setcopyright{cc}
\setcctype{by}
\acmConference[WWW '26] {Proceedings of the ACM Web Conference 2026}{April 13--17, 2026}{Dubai, United Arab Emirates.}
\acmBooktitle{Proceedings of the ACM Web Conference 2026 (WWW '26), April 13--17, 2026, Dubai, United Arab Emirates}
\acmISBN{979-8-4007-2307-0/2026/04}
\acmDOI{10.1145/3774904.3792701}
\usepackage{booktabs}
\usepackage{multirow}
\usepackage{enumitem}
\usepackage{bbding}
\usepackage{subfigure}
\setlist[itemize]{leftmargin=*, nosep}
\usepackage{siunitx}
\usepackage{hyperref}
\usepackage{color, xspace}
\usepackage{algorithm}
\usepackage{algorithmicx}
\usepackage{algpseudocode}
\usepackage{svg}

\newcommand{\name}{UrbanMoE\xspace} 
\newcommand{\etal}{\emph{et al.}\xspace} 
\newcommand{\eg}{\emph{e.g.,}\xspace} 
\newcommand{\ie}{\emph{i.e.,}\xspace} 

\newcommand{\moe}{Sparse Multi-Expert Module\xspace} 

\begin{document}

\title{UrbanMoE: A Sparse Multi-Modal Mixture-of-Experts Framework for Multi-Task Urban Region Profiling}



\author{Pingping Liu}
 \affiliation{%
   \institution{Jilin University}
   \city{Changchun}
  \country{China}
  }
\email{liupp@jlu.edu.cn}

 \author{Jiamiao Liu}
 \affiliation{%
   \institution{Jilin University}
   \city{Changchun}
  \country{China}
 }
\email{ljm24@mails.jlu.edu.cn}

 \author{Zijian Zhang}
 \authornote{Corresponding author.}
 \affiliation{
 \institution{Jilin University}
   \city{Changchun}
  \country{China}
  }
\email{zhangzijian@jlu.edu.cn}

 \author{Hao Miao}
 \affiliation{%
   \institution{The Hong Kong Polytechnic University}
   \city{Hong Kong}
   \country{China}
   }
\email{hao.miao@polyu.edu.hk}

 \author{Qi Jiang}
 \affiliation{ 
  \institution{Jilin University}
   \city{Changchun}
  \country{China}
  }
 \email{jiangqi5523@mails.jlu.edu.cn}

 \author{Qingliang	Li}
 \affiliation{
   \institution{Changchun Normal University}
   \city{Changchun}
  \country{China}
 }
 \email{liqingliang@ccsfu.edu.cn}

 \author{Qiuzhan	Zhou}
 \affiliation{
   \institution{Jilin University}
   \city{Changchun}
  \country{China}
  }
 \email{zhouqz@jlu.edu.cn}

 \author{Irwin	King}
 \affiliation{
   \institution{The Chinese University of Hong Kong}
   \city{Hong Kong}
  \country{China}
  }
 \email{king@cse.cuhk.edu.hk}

\renewcommand{\shortauthors}{Pingping Liu et al.}

\begin{abstract}
Urban region profiling, the task of characterizing geographical areas, is crucial for urban planning and resource allocation. 
However, existing research in this domain faces two significant limitations. 
First, most methods are confined to single-task prediction, failing to capture the interconnected, multi-faceted nature of urban environments where numerous indicators are deeply correlated. 
Second, the field lacks a standardized experimental benchmark, which severely impedes fair comparison and reproducible progress.
To address these challenges, we first establish a comprehensive benchmark for multi-task urban region profiling, featuring multi-modal features and a diverse set of strong baselines to ensure a fair and rigorous evaluation environment. 
Concurrently, we propose UrbanMoE, the first sparse multi-modal, multi-expert framework specifically architected to solve the multi-task challenge.
Leveraging a sparse Mixture-of-Experts architecture, it dynamically routes multi-modal features to specialized sub-networks, enabling the simultaneous prediction of diverse urban indicators.
We conduct extensive experiments on three real-world datasets within our benchmark, where UrbanMoE consistently demonstrates superior performance over all baselines. Further in-depth analysis validates the efficacy and efficiency of our approach, setting a new state-of-the-art and providing the community with a valuable tool for future research in urban analytics\footnote{The code is available at \url{https://github.com/JLU-LJM/UrbanMoE}}.



\end{abstract}





\maketitle

\section{INTRODUCTION}
Urban region profiling, which aims to understand and quantify the fine-grained characteristics of urban areas, is a cornerstone of modern urban computing. 
Accurate estimations of metrics such as population density~\cite{gebru2017using,he2020population,wang2022population}, economic vitality~\cite{he2018perceiving,liu2023knowledge,abitbol2020interpretable,yeh2020using,jean2016combining}, and environmental impact~\cite{lu2017predicting,naz2023comparative} are crucial for a wide range of applications, such as smart city planning and policy-making. 
{The rapid growth of diverse web- and IoT-sourced urban data, \eg satellite imagery, POIs, and sensor networks, offers unprecedented opportunities to capture both physical and functional characteristics of urban areas. 
Effectively integrating these multi-source data enables more holistic and accurate region profiling, supporting intelligent services that respond adaptively to urban dynamics~\cite{huang2023learning,wang2025multi,xi2022beyond}.
}


{
Recent research on urban region profiling has progressed from single-modality learning to multi-modal fusion.
Early works~\cite{yeh2020using,he2018perceiving, park2022learning,jean2019tile2vec} focused on extracting features from a single source, such as satellite imagery or POIs, to represent urban areas and predict socioeconomic indicators.
More recent studies~\cite{xi2022beyond,huang2021m3g,wang2025multi} have highlighted the importance of integrating heterogeneous data sources to obtain richer urban representations.
Inspired by advances in vision-language pretraining, some works~\cite{yan2024urbanclip,hao2025urbanvlp,xiao2024refound} further explore visual-language alignment for interpretable urban representations, 
while others~\cite{hao2025nature,liu2023knowledge} incorporate external knowledge and multi-level geospatial aggregation to enhance generalization. 
Together, these research efforts demonstrate a clear and consistent trend toward comprehensive, multimodal urban profiling. 
}

\begin{figure}[!t]
\setlength{\belowcaptionskip}{-6mm}
{{\includegraphics[width=0.9\linewidth]{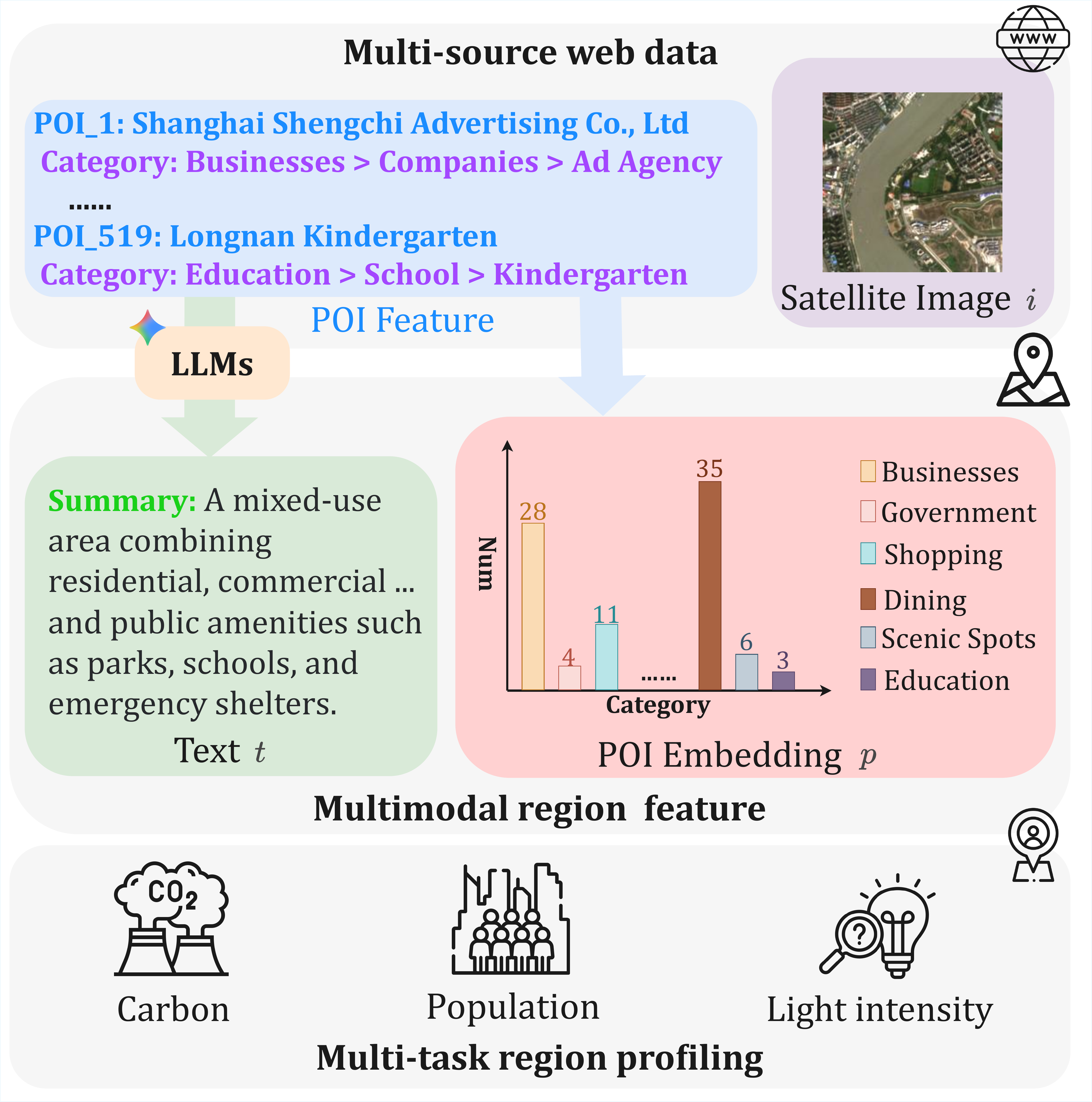}}}
\caption{Multimodal multi-task urban region profiling dataset construction.}
    \label{data}
\end{figure}

However, existing research in this domain faces two significant limitations. 
On one hand, the vast majority of current methods~\cite{hao2025urbanvlp,yan2024urbanclip,hao2025nature,xi2022beyond} focus on single-task prediction, such as PG-SimCLR or UrbanCLIP. 
This approach overlooks the practical needs of real-world scenarios where multiple metrics are often required simultaneously. 
More importantly, it neglects the inherent common patterns and complex interdependencies that exist among different urban tasks. 
For example, high population density and elevated carbon emissions are often geographically correlated. 
On the other hand, the field currently lacks a publicly available, well-aligned multi-modal benchmark for evaluating models on multi-modal urban profiling challenges. 
This absence of a standardized experimental environment hinders fair comparison and impedes reproducible progress in the community.
To address the aforementioned limitations, this paper aims to answer the following key research questions:
\textbf{Q1.} How can we design a unified framework to solve multi-task urban profiling by effectively capturing the shared patterns and complex relationships among different tasks?
\textbf{Q2.} How can the proposed framework achieve high performance without sacrificing computational efficiency, ensuring its practicality for large-scale urban applications?
\textbf{Q3.} How can we establish a fair and comprehensive benchmark for the rigorous evaluation and comparison of multi-modal, multi-task region profiling models?

Answering these questions presents several significant research challenges. 
First, the \textbf{inherent heterogeneity of multi-modal data} is a major hurdle. Urban data sources, such as structured POI statistics, unstructured text, and visual satellite image, reside in completely different feature spaces. 
Designing an effective fusion mechanism that can learn a shared, meaningful representation is non-trivial. 
Second, \textbf{effectively modeling inter-task relationships} is challenging. While tasks are related, naively sharing all model parameters can lead to "negative transfer," where learning one task degrades the performance of another. The key challenge lies in creating an architecture that enables selective knowledge sharing, capturing common patterns while preserving task-specific knowledge. 
Finally, there is a \textbf{fundamental trade-off between model performance and computational efficiency}. 
Sophisticated models capable of handling multi-modal and multi-task learning often come with prohibitively high computational costs, which significantly limit their practical deployment.

In this paper, we systematically address these challenges. Our work is summarized as follows:
\textbf{First}, to answer Q1 and Q2, we propose a novel sparse multi-modal Mixture-of-Experts framework for multi-task urban profiling, namely \name. 
It employs specialized expert networks to capture diverse data patterns and task correlations, while a sparse gating mechanism ensures that only the most relevant experts are activated for a given input, thus achieving a strong balance between performance and computational cost.
\textbf{Second}, to tackle the lack of a standardized evaluation platform (Q3), we construct and release a new benchmark featuring three diverse cities. 
To provide a holistic view of each region, the benchmark incorporates three distinct data modalities: 
(1) visual features from satellite images; 
(2) textural features generated by a Large Language Model (LLM) summarizing POI features; 
and (3) POI embeddings that represent the count of each POI category, indicating structural information. 
On this benchmark, we define three corresponding profiling tasks: carbon emission, population, and night light intensity.
{Figure~\ref{data} presents our benchmark's data processing pipeline, 
web data of POI feature and satellite image are collected 
demonstrating how these data sources are organized into a structured pipeline for urban region profiling.}
\textbf{Finally}, we conduct comprehensive experiments on our benchmark. 
The results verify that \name achieves consistently superior performance against a wide range of baselines. 
Our in-depth analyses, including extensive ablation studies and gate mechanism visualizations, further validate the efficacy, efficiency, and interpretability of our proposed framework.


The main contributions can be summarized as follows:
\begin{itemize}
    \item We establish a benchmark for region profiling, including multi-modal features, multi-task, and diverse baselines, to foster a fair and standardized experimental environment.
    \item For the first time, we propose a sparse multi-modal multi-experts framework, \ie \name, specifically designed to address multi-task urban region profiling.
    \item Extensive experiments on our benchmark with 3 real-world datasets verify the consistently superior performance of our proposed \name across various metrics. In-depth analysis further supports its efficacy and efficiency.
\end{itemize}

\section{METHODOLOGY}

\subsection{Problem Definition}
Given $K$ regions in a city, 
let $\mathcal{I} = \left \{ \boldsymbol{i}_{k} \right \}_{k=0}^{K}  $ denote the set of satellite images, $\mathcal{P} = \left \{\boldsymbol{p}_{k}    \right \}_{k=0}^{K} $ 
represent the POI-based features comprising attributes such as names and categories, and   $\mathcal{T} = \left \{ \boldsymbol{t}_{k} \right \}_{k=0}^{K} $ indicate the textual descriptions. 
The goal of \textbf{Multi-Task Urban Region Profiling} is to leverage these multimodal information, \ie $\left \{ \mathcal{I},\mathcal{P},\mathcal{T}
\right \} $, to accurately predict multiple urban indicators $\hat{\mathcal{Y}}$, \eg
carbon emissions $\boldsymbol{\hat{y}}_{c,k}$, population $\boldsymbol{\hat{y}}_{p,k}$, and nighttime light intensity $\boldsymbol{\hat{y}}_{l,k}$ for all regions. 
Figure~\ref{data} illustrates the data for a single region as an example, including its satellite imagery, POI Feature, generated textual description, and POI embedding, providing an informative multimodal input structure used for prediction.

\subsection{Framework Overview}

\begin{figure*}[!t]
\setlength{\belowcaptionskip}{-3mm}
{{\includegraphics[width=1\linewidth]{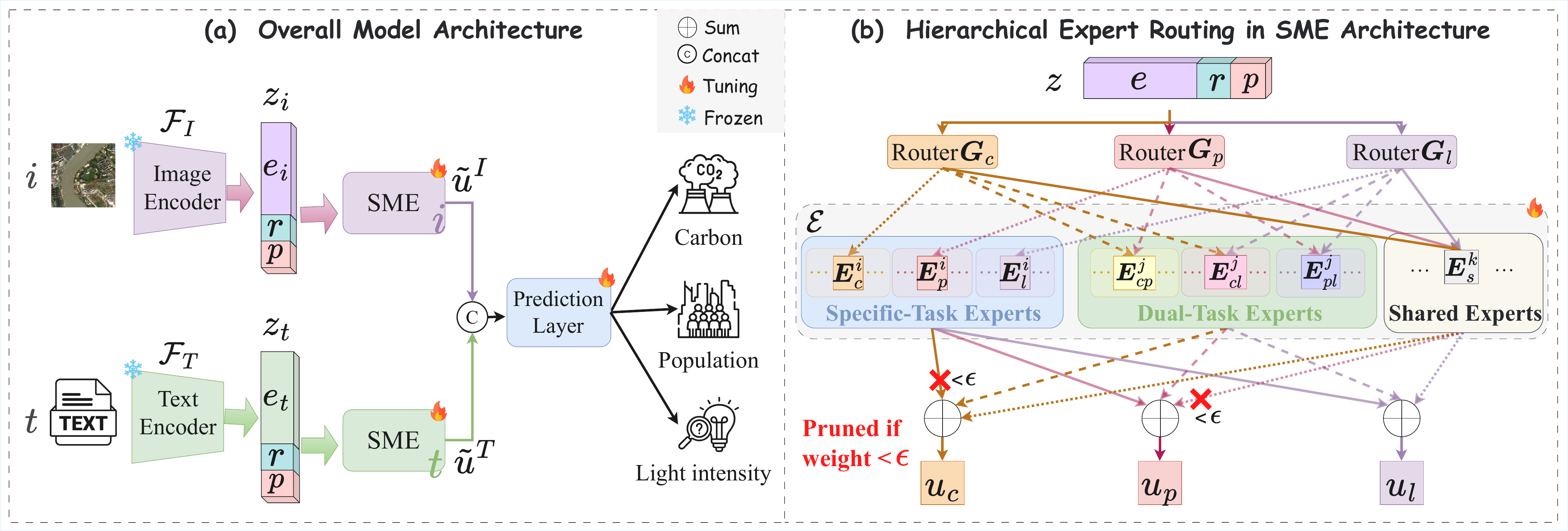}}}
\caption{The proposed \name framework, including (a) overall model architecture and (b) hierarchical expert routing.}
    \label{model}
\end{figure*}

Considering the intrinsic dependency among the diverse region profiling demands and informative region features from multiple sources, we build a multimodal sparse mixture-of-experts framework, \name, for multi-task urban region profiling.
On one hand, it is required to handle the intrinsic heterogeneity of urban data, which naturally arises from multiple data sources such as satellite imagery, textual descriptions, and POI distributions.
On the other hand, \name needs to balance the multiple optimization objectives to achieve the trade-off between information gain from multi-task joint patterns and the negative transfer.

Specifically, \name consists of three primary components, as illustrated in Figure~\ref{model} (a).
From left to right, \textit{Multimodal feature extraction layer} captures distinct modality-specific embeddings from the input image, text, and POI data. 
Dual-branch \textit{sparse multi-expert module} integrates multi-modal information and models inter-task dependencies through sparse expert networks.
\textit{Task-aware representation fusion} jointly estimates multiple urban indicators through dynamically weighted outputs.

\subsection{Multimodal Feature Extration}

Given the satellite image and POI attributes for a region, we construct its multimodal representation by extracting and integrating signals from these heterogeneous data sources.

\subsubsection{Image Representation ($\boldsymbol{e}_i$).}
The satellite image offers a visual snapshot of the region's physical environment, revealing crucial characteristics like land use, vegetation cover, and infrastructure. 
To capture these visual cues, which are fundamental to understanding a region's natural endowment and development stage, we leverage the powerful pre-trained model, RemoteCLIP~\cite{liu2024remoteclip}.
Its strong cross-modal alignment and zero-shot generalization capabilities on remote sensing imagery allow us to effectively process the satellite image $\boldsymbol{i}$ and obtain its corresponding visual representation $\boldsymbol{e}_i$ without any fine-tuning, \ie $\boldsymbol{e}_i = \mathcal{F}_I(\boldsymbol{i}) \in \mathbb{R}^{d_e}$, where $\mathcal{F}_I(\cdot)$ represents the pretrained visual encoder and $d_e$ is the embedding size.

\subsubsection{POI Embedding ($\boldsymbol{p}$).}
While many existing methods rely solely on satellite imagery~\cite{law2019take,he2018perceiving,jean2019tile2vec,park2022learning}, we argue that the distribution of POIs is a vital and complementary indicator for urban profiling. 
POI data directly quantifies the configuration of commercial, service, and public facilities, precisely delineating an area's functional role. 
This information is critical because two regions with similar physical appearances in satellite images may serve vastly different urban functions if their POI distributions diverge. 
Therefore, we systematically extract the number of POIs in each major category within a region and construct a statistical POI embedding $\boldsymbol{p}$ to represent its functional characteristics.


\subsubsection{Text Representation ($\boldsymbol{e}_t$).}
To capture a more abstract, high-level understanding of a region's character, we distill the informative POI features into an interpretable textual summary with Large Language Models. 
This summary provides a semantic description of the area's overall nature. To encode this information, we again utilize RemoteCLIP, feeding the textual description into its text encoder. 
This process yields a text representation $\boldsymbol{e}_t$ that captures the region's essential semantic characteristics, \ie $\boldsymbol{e}_t = \mathcal{F}_T(\boldsymbol{t}) \in \mathbb{R}^{d_e}$, where $\mathcal{F}_T(\cdot)$ is the textual encoder. In our case, we use RemoteCLIP as the encoder that simultaneously produces the visual and textual representations.


\subsubsection{Multimodal Representation Generation.}
\label{sec:MRG}



To enrich the modality-specific features with shared contextual priors, we introduce two learnable embeddings: a region embedding $\boldsymbol{r}\in \mathbb{R}^{d_r}$ designed to capture the unique, intrinsic characteristics of each urban area, and a POI embedding $\boldsymbol{p}\in \mathbb{R}^{d_p}$ that reflects its functional composition and the density of various amenities. 
These context-aware embeddings are then concatenated with both the image representation $\boldsymbol{e}_i$ and the text representation $\boldsymbol{e}_t$ to form the final input vectors for our two multimodal branches. Formally, this is expressed as:

\begin{equation}\label{Equ:zi_zt}
\mathbf{z}_i = [\mathbf{e}_i; \mathbf{r}; \mathbf{p}] \in \mathbb{R}^{d_{in}}, \\
\mathbf{z}_t = [\mathbf{e}_t; \mathbf{r}; \mathbf{p}] \in \mathbb{R}^{d_{in}}, \quad
d_{in} = d_e + d_r + d_p. 
\end{equation}

This fusion step is critical as it creates a unified feature space where modality-specific signals, \ie the visual cues from the image ($\boldsymbol{e}_i$) and the semantic content from the text ($\boldsymbol{e}_t$), are grounded by identification ($\boldsymbol{r}$) and functional ($\boldsymbol{p}$) context. 
By preserving the unique information from each modality while enriching it with these priors, we construct a comprehensive and robust representation of the urban environment, providing a solid foundation for the subsequent multi-task processing.

\subsection{\moe}
{
To address the challenges of effective multimodal representation learning, balanced optimization across heterogeneous urban indicators, and efficient modeling of large-scale datasets, we propose a dual-branch architecture. This design incorporates one branch for modeling image-based spatial attributes and another for encoding textual semantics from POI descriptions, thereby enabling modality-specialized processing and effective cross-modal integration.


Each branch is equipped with an identical Sparse Multi-Expert (SME) module. As shown in Figure \ref{model} (b), the SME refines modality-specific features through the \textit{multi-view expert networks}, \textit{sparse router}, and \textit{task-aware representation fusion}. For clarity, we take one SME as an example to elaborate on its design, as the same structure is applied symmetrically to both image and text branches. 
The refined outputs are then integrated across modalities, linking the dual-branch architecture with fine-grained multi-task learning and ultimately balancing efficiency, flexibility, and predictive accuracy in urban profiling. 


{
Compared with existing MoE methods~\cite{ma2018modeling} which include all experts' output as result, our \name selectively activates several most relevant experts to calculate, enabling leading efficiency.
}

\subsubsection{Multi-View Expert Networks}

{
} 

{
To capture diverse patterns across heterogeneous urban indicators, we design each SME with a multi-view expert structure. 
Specifically, the experts are organized into three complementary categories:
specific-task experts $\boldsymbol{E}_{sp} $, dual-task experts $\boldsymbol{E}_{dt} $, and shared experts $\boldsymbol{E}_{sh} $. 
}
This hierarchical design allows selective activation of task-specific signals while retaining cross-task generalization ability. Each expert is implemented as a lightweight feedforward network, formulated as:
\begin{equation}\label{Equ:Ez}
   \boldsymbol{E}(z)= \sigma(\boldsymbol{W}_2\phi (\boldsymbol{W}_1z+\boldsymbol{b}_1)+\boldsymbol{b}_2),
\end{equation}

{\noindent where $\boldsymbol{W}_1$ and $\boldsymbol{W}_2$ are learnable weight matrices, 
$\boldsymbol{b}_1$ and $\boldsymbol{b}_2$ are learnable bias terms, 
$\phi(\cdot)$ denotes a non-linear activation function, 
and $\sigma(\cdot)$ is an output normalization function.}

\noindent \textbf{Specific-Task Experts ($\boldsymbol{E}_{sp}$)}. 
{
To enable fine-grained specialization for each task, we introduce Specific-Task experts organized as $\boldsymbol{E}_{sp} = \{\boldsymbol{E}_c,\boldsymbol{E}_p,\boldsymbol{E}_l\} $, 
where $\boldsymbol{E}_c$, $\boldsymbol{E}_p$, and $\boldsymbol{E}_l$ correspond to the experts for carbon emission, population, and nighttime light intensity, respectively.
Formally, each task-specific expert set ($\boldsymbol{E}_c$, $\boldsymbol{E}_p$, $\boldsymbol{E}_l$) consists of $N_{sp}$ experts per modality.
For example, 
nighttime light intensity prediction leverages illumination distribution from the image branch and POI indicators of economic activity from the text branch. 
This dual-branch specialization enables each task to fully exploit complementary multimodal information.
}

\noindent \textbf{Dual-Task Experts ($\boldsymbol{E}_{dt}$)}. 
{
To capture dependencies between pairs of tasks, we introduce dual-task experts shared between task pairs, formally defined as  $\boldsymbol{E}_{dt} = \{\boldsymbol{E}_{cp},\boldsymbol{E}_{cl},\boldsymbol{E}_{pl}\}$, where $\boldsymbol{E}_{cp}$ links carbon and population, $\boldsymbol{E}_{cl}$ connects carbon and nighttime light, and $\boldsymbol{E}_{pl}$ associates population and nighttime light.
Formally, each dual-task expert set ($\boldsymbol{E}_{cp}$, $\boldsymbol{E}_{cl}$, $\boldsymbol{E}_{pl}$) consists of $N_{dt}$ experts per modality to capture cross-task patterns and complementary information.
Focusing on the joint modeling of three tasks, this module enables fine-grained, comprehensive multi-objective optimization.
}

\noindent \textbf{Shared Experts ($\boldsymbol{E}_{sh}$)}. 
{Task-agnostic shared experts ($\boldsymbol{E}_{sh}$) provide global knowledge across all tasks, capturing high-level representations.
Formally, $\boldsymbol{E}_{sh}$ consists of $N_{sh}$ experts per modality.
Image-branch shared experts focus on universal spatial cues (\eg road networks, land-use diversity, urban morphology), while text-branch shared experts encode functional and socio-economic semantics from POIs. 
These shared experts form the backbone of the SME framework, offering stable and generalizable features that enhance robustness and serve as a foundation for downstream predictions.}



After defining the three expert categories, the complete expert pool is formulated as:
\begin{equation}\label{E}
   \mathcal{E} = \boldsymbol{E}_{sp} \cup \boldsymbol{E}_{dt} \cup  \boldsymbol{E}_{sh} .
\end{equation}

{ This expert pool captures the intrinsic patterns for every task and models dependencies among tasks. 
}

\subsubsection{Sparse Router} 
{
We design task-specific router functions as a lightweight gating mechanism to determine how each input feature is dispatched to the SME expert pool.
Specifically, we implement three routers corresponding to three tasks, 
\ie carbon emissions ($\boldsymbol{G}_c$), population ($\boldsymbol{G}_p$), and nighttime light intensity ($\boldsymbol{G}_l$). 
We denote $\boldsymbol{G}_\tau \in \mathbb{R}^M $ as the gating weights over the $M$ experts for task ${\tau}$. 
}


Let $\boldsymbol{z}\in \mathbb{R} ^{d_{in}} $ denote the input feature vector, corresponding to the multimodal representation extracted in Section ~\ref{sec:MRG}, 
where $d_{in}$ is the dimensionality of the concatenated embeddings from image or text features together with the region and POI embeddings. The task-specific gating weights are computed via a linear transformation followed by a Softmax function:
\begin{equation}\label{Gz}
    \boldsymbol{G}_\tau(z) = \mathrm{Softmax}\left ( \boldsymbol{W}_\tau z +\boldsymbol{b}_\tau  \right ),
\end{equation}
\noindent where $\boldsymbol{W}_\tau$ and $\boldsymbol{b}_\tau$ are the learnable weight and bias for task $\tau$. 

To promote computational efficiency and feature sparsity, a threshold $\epsilon$ is applied to the gating weights. Only experts with $\boldsymbol{G}_\tau > \epsilon$ are activated, while the others are set to zero:

\begin{equation}\label{Ge}
    \tilde{\boldsymbol{G}_\tau} =
    \begin{cases}
    \boldsymbol{G}_\tau, & \text{if } \boldsymbol{G}_\tau > \epsilon \\
    0, & \text{otherwise}.
    \end{cases}
\end{equation}

{This threshold-based sparse routing mechanism offers key advantages.
By activating only the most relevant experts for each input, it reduces computational and storage overhead. 
{The router retains only the most informative gating channels, effectively performing feature selection along the expert dimension and concentrating input representations on task-relevant information.} 
This mechanism also prevents less informative experts from impairing prediction quality, while adaptively routing inputs to the optimal expert subset based on semantic content. 
Consequently, it achieves task-specific specialization while preserving cross-task knowledge, effectively balancing efficiency, flexibility, and accuracy within the multimodal framework.}

\subsubsection{Task-Aware Representation Fusion}
The final stage of the SME framework is the task-specific output layer, which integrates the routing weights $\boldsymbol{G}$ with the expert network $\mathcal{E}$ to generate task-dependent representations.

Given the expert pool $\mathcal{E}_\tau$, , the output representation for task $\tau$ is computed as a weighted combination of the activated experts:
\begin{equation}\label{add}
    \tilde{\boldsymbol{u}}_\tau = \sum_{m=1}^{M} \tilde{\boldsymbol{G}}_{\tau,m}\boldsymbol{E}_m (z),  
\end{equation}

\noindent where $\boldsymbol{z}$ is the multimodal input representation, $\boldsymbol{E}_m(\cdot)$ denotes the transformation performed by the $m$-th expert, and $\boldsymbol{G}_{\tau,m}$ is the gating weight assigned to that expert for task $\tau$. This mechanism allows each task representation to be dynamically composed of specialized, dual-task, and shared knowledge sources, ensuring both precision and generalization.

{
In summary, the SME module implements a hierarchical sparse mixture-of-experts architecture to integrate multimodal signals. 
On the one hand, it provides an interpretable framework for modeling complex dependencies across heterogeneous urban tasks. 
On the other hand, task-aware specialization and adaptive sparse routing enable selective expert activation, ensuring computational efficiency while capturing both task-specific and cross-task patterns that existing mixture-of-experts methods overlook.
}

\subsection{Prediction and Objective Function}

After feature extraction and expert routing, the outputs from the dual-branch \moe modules are integrated to form task-specific multimodal representations.
Specifically, for each task $\tau$ , the image-branch output $\tilde{\boldsymbol{u}}_\tau^I$ and the text-branch output $\tilde{\boldsymbol{u}}_\tau^T$ are concatenated and directly fed into a task-specific prediction layer, yielding:

\begin{equation}\label{MLP}
   \hat{\boldsymbol{y}} _\tau  = \mathrm{MLP}_\tau (\tilde{\boldsymbol{u}}_\tau^I \parallel \tilde{\boldsymbol{u}}_\tau^T).
\end{equation}

Collectively, the model produces: $\mathcal{\hat{Y} } = \left \{ \boldsymbol{\hat{y}}_c,\boldsymbol{\hat{y}}_p,\boldsymbol{\hat{y}}_l \right \}$, corresponding to carbon emissions, population, and nighttime light intensity. 

For model training, we adopt a multi-task regression objective based on the Mean Squared Error (MSE). Let the ground-truth labels be $\mathcal{Y } = \left \{ \boldsymbol{y}_c,\boldsymbol{y}_p,\boldsymbol{y}_l \right \}$. The overall loss function is defined as:
\begin{equation}\label{LOSS}
   \mathcal{L}=\lambda_{c}\left\|\hat{\boldsymbol{y}}_{c}-\boldsymbol{y}_{c}\right\|_{2}^{2}+\lambda_{p}\left\|\hat{\boldsymbol{y}}_{p}-\boldsymbol{y}_{p}\right\|_{2}^{2}+\lambda_{l}\left\|\hat{\boldsymbol{y}}_{l}-\boldsymbol{y}_{l}\right\|_{2}^{2},
\end{equation}
\noindent where $\lambda_{c}$, $\lambda_{p}$, $\lambda_{l}$ are weighting coefficients balancing the contributions of each task.


\subsection{Discussion}
{
Unlike traditional multi-task methods~\cite{jacobs1991adaptive}, we elaborate comprehensively fine-grained experts to model task dependencies explicitly. 
The structure, combined with sparse routing, allows the model to adaptively capture both specialized features and cross-task correlations, enhancing both performance and interpretability.
} For example, on BJ dataset, with $N_{sp}=8$, $N_{dt}=2$, and $N_{sh}=4$, the total expert pool contains $8\times 3+2\times 3+4=34$ experts per modality.  
{Take carbon emission as an example,
with the optimal threshold $\epsilon=0.01$, the model sparsely activates $6+2+2+3=13$ experts selected from $\{\boldsymbol{E}_c , \boldsymbol{E}_{cp} , \boldsymbol{E}_{cl} , \boldsymbol{E}_{sh} \}$.
}
This design improves computational efficiency while maintaining strong representational capacity, providing a feasible solution for large-scale urban computing tasks.

\begin{table}[!t]
\renewcommand{\arraystretch}{1}
\centering
\setlength{\abovecaptionskip}{0.1mm}
\caption{Multimodal urban region profiling dataset statistics.}
\label{dataset}
\resizebox{0.48\textwidth}{!}{
\begin{tabular}{lccccc}
\toprule
\multirow{2}{*}{\textbf{Dataset}} & \multicolumn{2}{c}{\textbf{Coverage}} & \multirow{2}{*}{\textbf{\begin{tabular}[c]{@{}c@{}}\#Satellite\\ Image\end{tabular}}} & \multirow{2}{*}{\textbf{\begin{tabular}[c]{@{}c@{}}\#POI \end{tabular}}}& \multirow{2}{*}{\textbf{\begin{tabular}[c]{@{}c@{}}\#Task \end{tabular}}} \\
\cmidrule(lr){2-3} 
 & \textbf{Bottom-left} & \textbf{Top-right} & \\
\midrule
SH & 30.70$^{\circ}$N, 120.90$^{\circ}$E & 31.63$^{\circ}$N, 121.95$^{\circ}$E & \num{1558} & \num{1100075} & \num{3} \\
BJ & 39.65$^{\circ}$N, 115.82$^{\circ}$E & 40.65$^{\circ}$N, 117.11$^{\circ}$E & \num{2310} & \num{541407} & \num{3} \\
DC & 40.57$^{\circ}$N, 73.82$^{\circ}$W & 41.54$^{\circ}$N, 72.50$^{\circ}$W & \num{1091} & \num{164942} & \num{3} \\
\bottomrule
\end{tabular}}
\vspace{-6mm}
\end{table}

\section{EXPERIMENTS} 
In this section, we present the experimental results of \name, including comprehensive comparisons with diverse state-of-the-art baselines, visualization of expert gating weights, and ablation studies to verify the contribution of different expert components.  The hyperparameter analysis is provided in the \textbf{Appendix ~\ref{sec:Hyper-Parameter}}.

\subsection{Dataset}
Since there is currently no publicly available benchmark dataset for multimodal multi-task urban profiling, we construct datasets by collecting and aligning heterogeneous data sources across three representative cities: Shanghai (\textbf{SH}), Beijing (\textbf{BJ}), and Washington D.C. (\textbf{DC}). 
Satellite images are obtained from the Copernicus Data Space Ecosystem (Sentinel-2)~\cite{ecosystem2023copernicus}, with a fixed patch size of 200×200 pixels and a spatial resolution of 10 meters per pixel, corresponding to approximately 4 km$^2$ per patch. 
Textual information is generated from the corresponding POI data following the preprocessing pipeline of~\cite{wang2024urbandatalayer}, including key attributes such as POI names and categories, and enhanced into coherent, semantically rich regional descriptions using Gemini Flash 2.0~\cite{team2023gemini}.
In addition, structured POI embeddings are constructed by counting POIs across major functional categories. 
Urban indicators are collected from multiple sources: carbon emissions from ODIAC~\cite{oda2018open} as an environmental indicator, population density from WorldPop~\cite{tatem2017worldpop} as a social indicator, and nighttime light intensity from VIIRS DNB~\cite{elvidge2021annual} as a proxy for economic activity. 
Table~\ref{dataset} summarizes the data statistics, including the number of satellite images and POIs, as well as the coverage area. We randomly partition the dataset into 60\% for training, 20\% for validation, and 20\% for testing to ensure reliable evaluation.
Detailed dataset information, including data sample and LLM prompt, can be referred to \textbf{Appendix~\ref{sec:LLM prompt}}. 



\subsection{Baseline}

{To comprehensively position our proposed method in urban region profiling, we compare it with representative baselines from two lines: (a)\textbf{ urban region profiling}, including Autoencoder ~\cite{kramer1991nonlinear}, PCA ~\cite{tipping1999probabilistic}, ResNet-18 ~\cite{he2016deep}, Tile2Vec ~\cite{jean2019tile2vec}, READ ~\cite{han2020lightweight}, PG-SimCLR ~\cite{xi2022beyond}, UrbanCLIP ~\cite{yan2024urbanclip} and RemoteCLIP ~\cite{liu2024remoteclip}; and (b) \textbf{multi-task urban region profiling}, including UrbanCLIP ~\cite{yan2024urbanclip} and RemoteCLIP ~\cite{liu2024remoteclip}. Detailed baseline descriptions are provided in the Appendix ~\ref{baseline}.}

\begin{table*}[t!]
\renewcommand{\arraystretch}{1.2}
\centering
\setlength{\abovecaptionskip}{1mm}
\caption{Overall experimental results on SH, BJ, DC datasets. The best results are {bold} and the second-best are {underlined}.
}
\label{SH}
\scalebox{0.73}{
\begin{tabular}{ccccccccccccccccc}
\toprule
\multirow{2}{*}{\textbf{Dateset}} & \multirow{2}{*}{\textbf{Model}} & \multicolumn{3}{c}{\textbf{Carbon}} & \multicolumn{3}{c}{\textbf{Population}} & \multicolumn{3}{c}{\textbf{Light intensity}} & \multicolumn{3}{c}{\textbf{AVG}} & \multirow{2}{*}{\textbf{\begin{tabular}[c]{@{}c@{}}Params.\\(M)\end{tabular}}} & \multirow{2}{*}{\textbf{\begin{tabular}[c]{@{}c@{}}Max Mem.\\ (MB)\end{tabular}}} & \multirow{2}{*}{\textbf{\begin{tabular}[c]{@{}c@{}}Total\\ Time (s)\end{tabular}}} \\
\cmidrule(lr){3-5} \cmidrule(lr){6-8} \cmidrule(lr){9-11} \cmidrule(lr){12-14}
& &R$^2\mathcal{\uparrow }$ & RMSE$\mathcal{\downarrow}$ & MAE$\mathcal{\downarrow}$ & R$^2\mathcal{\uparrow }$ & RMSE$\mathcal{\downarrow}$ & MAE$\mathcal{\downarrow}$ & R$^2\mathcal{\uparrow }$ & RMSE$\mathcal{\downarrow}$ & MAE$\mathcal{\downarrow}$ & R$^2\mathcal{\uparrow }$ & RMSE$\mathcal{\downarrow}$ & MAE$\mathcal{\downarrow}$ & & & \\ 
\midrule
\multirow{13}{*}{\textbf{SH}} & Autoencoder   & -0.0201 & 3.1205 & 1.7086 & -0.0257 & 3.6487 & 1.6128 & -0.0575 & 0.7349 & 0.5441 & -0.0344 & 2.5014 & 1.2885 & 0.42 & \num{5980} & \num{381} \\
&PCA           & 0.2469  & 2.6812 & 1.2220 & 0.2432  & 3.1349 & 1.2456 & 0.2184  & 0.6318 & 0.4274 & 0.2362  & 2.1493 & 0.9650 & 0.50 & \num{10.3} & \num{411} \\
&ResNet-18     & 0.6700  & 1.7747 & 1.1049 & 0.5326  & 2.2943 & 1.1413 & 0.7189  & 0.4089 & 0.2575 & 0.6405  & 1.4926 & 0.8346 & 11.57 & \num{1629} & \num{1276} \\
&Tile2Vec      & 0.4950  & 2.4340 & 1.3480 & 0.4890  & 2.6620 & 1.2340 & 0.6160  & 0.4790 & 0.3210 & 0.5333  & 1.8583 & 0.9677 & 22.87 & \num{2160} & \num{5542} \\
&READ          & 0.2576  & 2.6505 & 1.2109 & 0.4426  & 2.7585 & 1.2624 & 0.5078  & 0.5292 & 0.3184 & 0.4027  & 1.9794 & 0.9306 & 22.36 & \num{2348} & \num{924} \\
&PG-SimCLR     & 0.7088  & 1.7176 & 1.0392 & 0.5352  & 2.2849 & 1.1136 & 0.5384  & 0.5215 & 0.3151 & 0.5941  & 1.5080 & 0.8226 & 11.47 & \num{5700} & \num{2899} \\
&UrbanCLIP (ST) & 0.7051  & 1.7189 & 1.0366 & 0.6710  & 2.0624 & 1.0691 & 0.7742  & 0.3723 & 0.2415 & 0.7168  & 1.3845 & 0.7824 & 639.04 & \num{12764} & \num{30758} \\
&RemoteCLIP (ST)& \underline{0.7529} & \underline{1.5460} & \underline{0.9406} & 0.6724 & 2.0144 & 1.0929 & \underline{0.7877} & \underline{0.3611} & \underline{0.2390} & 0.7377 & {1.3072} & {0.7575} & 430.65 & --- & \num{850962} \\

\cmidrule(lr){2-17}
&UrbanCLIP (MT) & 0.6835  & 1.7499 & 1.0374 & 0.6595  & 2.0982 & 1.1026 & 0.7559  & 0.3872 & 0.2398 & 0.6996  & 1.4118 & 0.7933 & 638.65 & \num{12764} & \num{20467} \\
&RemoteCLIP (MT)& 0.7450  & 1.5708 & 0.9807 & {0.6941} & {1.9887} & {1.0659} & 0.7809 & 0.3668 & 0.2409 & {0.7400} & 1.3088 & 0.7625 & 430.20 & --- & \num{843948}  \\
\cmidrule(lr){2-17}

&\name (ST) & 0.7387 &	1.5901	&0.8641 & \underline{0.7522} &	\underline{1.7900} &	\underline{0.8844} &	0.7792 &	0.3682 &	0.2415 &	\underline{0.7567} &	\underline{1.2494}	 &\underline{0.6633} &	7.75 &	\num{6840} &	\num{3982} \\

&\textbf{\name}          & \textbf{0.7837} & \textbf{1.4465} & \textbf{0.7953} & \textbf{0.7872} & \textbf{1.6585} & \textbf{0.7775} & \textbf{0.8001} & \textbf{0.3504} & \textbf{0.2320} & \textbf{0.7903} & \textbf{1.1518} & \textbf{0.6016} & 5.50 & \num{6894} & \num{1595} \\
\cmidrule(lr){2-17}
&Improvement   & 4.09\% & 6.44\% & 15.45\% & 13.41\% & 16.60\% & 37.09\% & 1.57\% & 2.96\% & 2.93\% & 6.80\% & 11.89\% & 20.58\% & --- & --- & --- \\
\bottomrule

\multirow{13}{*}{\textbf{BJ}} & Autoencoder & 0.1927 & 5.7791 & 1.7410 & 0.3789 & 0.7523 & 0.3889 & 0.4030 & 2.1651 & 1.1807 & 0.3249 & 2.8988 & 1.1035 & 0.42 & 5,980 & 430 \\
&PCA & 0.4738 & 4.5511 & 1.6064 & 0.4799 & 0.6884 & 0.3617 & 0.5676 & 1.8425 & 0.9973 & 0.5071 & 2.3607 & 0.9885 & 0.63 & 13.3 & 434 \\
&ResNet-18 & 0.6329 & 2.5737 & 1.3498 & 0.6847 & 0.4959 & 0.3069 & 0.7569 & 1.4601 & 0.7308 & 0.6915 & 1.5099 & 0.7958 & 11.57 & 1,629 & 1,543 \\
&Tile2Vec & 0.3700 & 5.5980 & 1.7040 & 0.6920 & 0.4980 & 0.3150 & 0.7230 & 1.5530 & 0.8300 & 0.5950 & 2.5497 & 0.9497 & 22.87 & 2,160 & 8,260 \\
&READ & 0.1204 & 6.3948 & 1.8325 & 0.6207 & 0.5537 & 0.3097 & 0.5865 & 1.7322 & 0.9808 & 0.4425 & 2.8936 & 1.0410 & 22.36 & 3,576 & 1,483 \\
&PG-SimCLR & 0.5538 & 3.7443 & 1.5579 & 0.6868 & 0.4931 & 0.3009 & 0.7882 & 1.3708 & 0.7148 & 0.6763 & 1.8694 & 0.8579 & 11.47 & 5,700 & 4,387 \\
&UrbanCLIP (ST) & {0.6661} & {2.1547} & {1.3196} & \textbf{0.8700} & \textbf{0.3299} & \textbf{0.1809} & 0.8141 & 1.1568 & 0.7144 & {0.7834} & 1.2138 & {0.7383} & 639.04 & 12,764 & 40,555 \\
&RemoteCLIP (ST) & 0.6560 & 2.1867 & 1.4413 & 0.8568 & 0.3465 & 0.2054 & 0.8349 & 1.0863 & 0.6353 & 0.7826 & 1.2065 & 0.7607 & 430.65 & --- & \num{855171} \\

\cmidrule(lr){2-17}
&UrbanCLIP (MT) & 0.6553 & 2.2584 & 1.4101 & 0.8599 & 0.3426 & 0.2121 & 0.8156 & 1.1520 & 0.6889 & 0.7769 & 1.2510 & 0.7704 & 638.65 & 12,764 & 31,043 \\
&RemoteCLIP (MT) & 0.5896 & 2.3887 & 1.5818 & 0.8557 & 0.3477 & 0.2237 & {0.8424} & {1.0649} & {0.6227} & 0.7626 & 1.2671 & 0.8094 & 430.20 & --- & \num{845317} \\
\cmidrule(lr){2-17}

&\name (ST) & \underline{0.7772} &	\underline{1.7599} &	\underline{0.9985} &	0.8624 &	0.3396 &	0.1864 &	\underline{0.8513} &	\underline{1.0345} &	\underline{0.5924} &	\underline{0.8303} &	\underline{1.0447} &	\underline{0.5924} &	7.75 &	\num{6840} &	\num{6279} \\

&\textbf{\name} & \textbf{0.7798} & \textbf{1.7499} & \textbf{0.9585} & \underline{0.8625} & \underline{0.3334} & \underline{0.1822} & \textbf{0.8642} & \textbf{0.9888} & \textbf{0.5602} & \textbf{0.8355} & \textbf{1.0240} & \textbf{0.5670} & 5.50 & 6,894 & 2,237 \\
\cmidrule(lr){2-17}
&Improvement & 17.07\% & 18.79\% & 27.36\% & -0.86\% & -1.06\% & -0.72\% & 2.59\% & 7.15\% & 10.04\% & 6.65\% & 15.13\% & 23.20\% & --- & --- & --- \\

\bottomrule

\multirow{13}{*}{\textbf{DC}} & Autoencoder & 0.0707 & 8.0247 & 3.2522 & 0.4096 & 4.0687 & 2.5430 & 0.4907 & 5.7097 & 3.6309 & 0.3237 & 5.9344 & 3.1420 & 0.42 & 5,980 & 360 \\
&PCA & 0.0132 & 8.2084 & 3.4326 & 0.2954 & 4.4446 & 2.6895 & 0.5421 & 5.4136 & 3.4294 & 0.2836 & 6.0222 & 3.1838 & 0.37 & 7.3 & 346 \\
&ResNet-18 & 0.3775 & 6.0623 & 2.7628 & 0.6730 & 3.6868 & 1.7222 & 0.6927 & 4.4645 & 2.6935 & 0.5811 & 4.7379 & 2.3928 & 11.57 & 1,629 & 1,005 \\
&Tile2Vec & 0.3090 & 6.3810 & 3.0250 & 0.5720 & 3.8870 & 1.9560 & 0.7260 & 4.3980 & 2.8480 & 0.5357 & 4.8887 & 2.6097 & 22.87 & 2,160 & 3,895 \\
&READ & 0.1971 & 7.8527 & 2.5832 & 0.6648 & 3.6670 & 1.8453 & 0.7637 & 4.0380 & 2.4045 & 0.5419 & 5.1859 & 2.2777 & 22.36 & 3,576 & 676 \\
&PG-SimCLR & 0.2378 & 7.0331 & 2.3510 & 0.4846 & 4.0018 & 1.7965 & 0.5608 & 5.3782 & 2.7677 & 0.4277 & 5.4710 & 2.3051 & 11.47 & 5,700 & 2,337 \\
&UrbanCLIP (ST) & \underline{0.6956} & \underline{2.8611} & {2.0451} &{0.7190} & {2.9297} & {1.4991} & 0.8597 & 3.1097 & \underline{1.8068} & {0.7581} & {2.9407} & {1.7837} & 639.04 & 12,764 & 26,117 \\
&RemoteCLIP (ST) & 0.5763 & 3.3756 & 2.3245 & 0.6979 & 3.0759 & 1.6562 & \underline{0.8627} & \underline{3.0648} & 2.0230 & 0.7123 & 3.1721 & 2.0012 & 430.65 & --- &\num{842993} \\

\cmidrule(lr){2-17}

&UrbanCLIP (MT) & 0.6830 & 2.9196 & 2.0620 & 0.7168 & 2.9782 & 1.5454 & 0.8502 & 3.2008 & 1.8868 & 0.7500 & 3.0329 & 1.8314 & 638.65 & 12,764 & 17,945 \\

&RemoteCLIP (MT) & 0.4610 & 3.8074 & 2.5414 & 0.6923 & 3.1042 & 1.6926 & 0.8607 & 3.0884 & 1.9831 & 0.6713 & 3.3167 & 2.0724 & 430.20 & --- & \num{849409} \\
\cmidrule(lr){2-17}

&\name (ST) & 0.6836&	2.9171&	 \underline{2.0424}&	\underline{0.8655}&	\underline{2.0519}&	\underline{1.1860}&	0.8601&	3.0892&	1.8264&	\underline{0.8031}& \underline{2.6861}&	\underline{1.6849}&	7.75&	\num{6840}&	\num{2892}
 \\

&\textbf{\name} & \textbf{0.6971} & \textbf{2.8567} & \textbf{2.0392} & \textbf{0.8775} & \textbf{1.9585} & \textbf{1.1726} & \textbf{0.8687} & \textbf{2.9962} & \textbf{1.7237} & \textbf{0.8144} & \textbf{2.6038} & \textbf{1.6452} & 5.50 & 6,894 & 1,024 \\
\cmidrule(lr){2-17}
&Improvement & 0.22\% & 0.15\% & 0.29\% & 22.04\% & 33.15\% & 21.78\% & 0.70\% & 2.24\% & 4.60\% & 7.43\% & 11.46\% & 7.76\% & --- & --- & --- \\
\bottomrule

\end{tabular}}
\end{table*}

\subsection{Experimental Setups}
To evaluate model performance, we use three standard regression metrics: the Coefficient of Determination (R$^2$), Root Mean Squared Error (RMSE), and Mean Absolute Error (MAE). 
To ensure the robustness and reproducibility of our results, all reported scores are averaged over three independent runs, each with a different random seed. 
For model training, we set the dimension of the POI-Embedding to $d_p=15$ and the Region-Embedding to $d_r=6$. 
The learning rate is fixed at a constant value of $3\times10^{-4}$ throughout all experiments.
In the expert configuration, the framework supports three prediction tasks (carbon emissions, population, and nighttime light intensity). The expert pool consists of $N_{sp}=8$ task-specific experts, $N_{dt}=2$ dual-task experts, and $N_{sh}=4$ shared experts. All experts share the same lightweight feed-forward architecture.

{To isolate the benefits of multi-task learning, we also evaluate UrbanMoE in a single-task (ST) setting, \ie UrbanMoE (ST), where all other settings remain consistent with the original version. 
In this configuration, each run focuses on a single task, with 16 experts and a dedicated output head. 
UrbanCLIP and RemoteCLIP are state-of-the-art methods for single-task urban region profiling, \ie UrbanCLIP (ST) and RemoteCLIP (ST).
To evaluate the performance of these state-of-the-art models in multi-task setting, we modify their output layers to produce three-dimensional predictions, corresponding to carbon emissions, population, and nighttime light intensity, \ie UrbanCLIP (MT) and RemoteCLIP (MT).
}

\begin{table*}[!t]
\renewcommand{\arraystretch}{1.2}
\centering
\setlength{\abovecaptionskip}{1mm}
\caption{Ablation study results with different expert configurations.}
\label{Expert}
\scalebox{0.75}{

\begin{tabular}{ccccccccccccccc}
\toprule
\multicolumn{3}{c}{\textbf{Types of Experts}}     & \multicolumn{3}{c}{\textbf{Carbon}}                 & \multicolumn{3}{c}{\textbf{Population}}             & \multicolumn{3}{c}{\textbf{Light intensity}}        & \multicolumn{3}{c}{\textbf{AVG}}                    \\
\cmidrule(lr){1-3} \cmidrule(lr){4-6} \cmidrule(lr){7-9} \cmidrule(lr){10-12}
\cmidrule(lr){13-15}
\textbf{Specific} & \textbf{Shared} & \textbf{Dual} & R$^2$↑             & RMSE↓           & MAE↓            & R$^2$↑             & RMSE↓           & MAE↓            & R$^2$↑             & RMSE↓           & MAE↓            & R$^2$↑             & RMSE↓           & MAE↓            \\ \midrule
 \Checkmark      & \XSolidBrush      & \XSolidBrush    & 0.7637          & 1.5122          & 0.8467          & 0.6724          & 2.0579          & 0.9247          & \underline{0.7844}    & \underline{0.3639}    & 0.2346          & 0.7402          & 1.3113          & 0.6687          \\
\XSolidBrush       & \Checkmark      & \XSolidBrush    & 0.7647          & 1.5087          & 0.8230          & 0.7546          & 1.7812          & 0.7988          & 0.7746          & 0.3721          & 0.2446          & 0.7646          & 1.2207          & \underline{0.6221}    \\
\XSolidBrush       & \XSolidBrush      & \Checkmark    & 0.7819          & 1.4509          & \underline{0.8082}    & 0.7233          & 1.8912          & 0.8449          & 0.7537          & 0.3890          & 0.2374          & 0.7530          & 1.2437          & 0.6302          \\
\Checkmark      & \Checkmark      & \XSolidBrush    & 0.7599          & 1.5240          & 0.8273          & \underline{0.7695}    & \underline{ 1.7263}    & 0.8322          & 0.7680          & 0.3775          & \underline{0.2333}    & 0.7658          & 1.2093          & 0.6309          \\
\Checkmark       & \XSolidBrush      & \Checkmark    & \textbf{0.7859} & \textbf{1.4417} & 0.8135          & 0.7564          & 1.7745          & \underline{0.8010}    & 0.7647          & 0.3801          & 0.2529          & \underline{0.7690}    & \underline{1.1988}    & 0.6225          \\
\XSolidBrush       & \Checkmark      & \Checkmark    & 0.7800          & 1.4591          & 0.8158          & 0.7097          & 1.9373          & 0.8406          & 0.7588          & 0.3849          & 0.2537          & 0.7495          & 1.2604          & 0.6367          \\
\Checkmark      & \Checkmark     & \Checkmark   & \underline{0.7837}    & \underline{1.4465}    & \textbf{0.7953} & \textbf{0.7872} & \textbf{1.6585} & \textbf{0.7775} & \textbf{0.8001} & \textbf{0.3504} & \textbf{0.2320} & \textbf{0.7903} & \textbf{1.1518} & \textbf{0.6016} \\ \bottomrule
\end{tabular}}
\vspace{-3mm}
\end{table*}

\subsection{Overall Performance}
An empirical evaluation of different models is performed on the three datasets.
As summarized in Table ~\ref{SH}, 
which correspond to SH, BJ, and DC, respectively,
the experimental results lead to the following observations:
\textbf{\textit{(1)}} 
The multi-task (MT) versions of UrbanCLIP and RemoteCLIP occasionally outperform their single-task (ST) counterparts, \eg Population prediction of RemoteCLIP on SH. 
It indicates the existence of transferable knowledge across tasks that joint optimization can leverage. 
However, the observed seesaw effect also reveals the complexity of task relationships and limitations in current multi-task balancing mechanisms.
\textbf{\textit{(2)}} 
Our \name demonstrates robust performance under the ST setting across multiple metrics, \eg Population prediction on SH and DC,  confirming its ability to effectively model multimodal urban characteristics.
Moreover, in the MT setting, it consistently outperforms its ST counterpart across all tasks, demonstrating successful cross-task knowledge transfer while effectively alleviating the seesaw phenomenon.
\textbf{\textit{(3)}} 
{
Our \name consistently outperforms the best-performing baseline on all datasets, achieving an average improvement of 10.70\% R$^2$ for population prediction, 6.89\% for carbon emission, and 1.61\% for light intensity.
These heterogeneous performance gains indicate that different tasks benefit from multi-task joint training in distinct ways, depending on their characteristics and reliance on shared representations. 
The significant improvement in population prediction demonstrates our model's ability to capture inter-task relationships, particularly those related to demographic patterns, while the gains in carbon emission and light intensity indicate its effectiveness across diverse urban prediction tasks.
} 
\textbf{\textit{(4)}} 
An exception to \name's superior performance is the population task on the BJ dataset, where our method slightly trails UrbanCLIP by a narrow margin (within 2\%).
This is primarily due to the characteristics of multi-task optimization: 
Improving population prediction provides substantial informational gains for the other two tasks, which in turn constrain the population-specific information in the region representation. 
Nevertheless, it is noteworthy that our method still outperforms the UrbanMoE single-task (ST) version across all indicators, indicating that each task benefits from joint training.
Moreover, our approach achieves the highest average scores overall, with an MAE improvement of over 20\%, demonstrating that UrbanMoE effectively reaches a globally optimal trade-off among heterogeneous urban tasks.
\textbf{\textit{(5)}} 
{
Our model achieves state-of-the-art performance while maintaining leading efficiency in both computation and memory usage.
Specifically, compared with UrbanCLIP, it reduces parameter count by over 99\% (5.5M vs. 638M) and memory usage to only 6.8GB, which can be ascribed to frozen pretrained encoders and sparse expert activation.
Moreover, our model converges rapidly, as shown in the remarkably shorter total time. 
This demonstrates the high efficiency of our approach for large-scale, multi-task region profiling. 
Note that the efficiency metrics of RemoteCLIP are taken from its original paper, and the maximum memory metric is unavailable, \ie denoted as --.
}    

\subsection{Ablation Study}
\label{sec:Effect of Expert Types}
{To further investigate the contributions of specific components, we design two dedicated studies: 
\textit{Effect of Expert Types}, which examines the impact of different expert configurations within the SME module, and 
\textit{Effect of Embedding}, which analyzes the influence of region and POI embeddings on multimodal representation learning and downstream task performance. 
All ablation studies are conducted on the SH dataset.
}

\noindent \textbf{Effect of Expert Types}. 
{
In this ablation study, each task is allocated a total of 16 experts (allocation details in \textbf{Appendix ~\ref{Expert Allocation}}). Based on the results in Table~\ref{Expert}, several conclusions can be drawn:
\textit{(1)}
In most cases, specific-task, dual-task, and shared experts all contribute stably to multi-task predictions, demonstrating that the SME module effectively integrates complementary information from different expert types to support robust joint learning across tasks.
\textit{(2)} 
Among the expert types, dual-task experts are particularly important for carbon emission prediction, consistently achieving $R^2 > 0.78$.
This reflects strong correlations between carbon and other tasks that the model can exploit, while population and light intensity also gain partial benefits from cross-task signals.
\textit{(3)}
Specific-task experts play a major role in light intensity prediction, capturing dedicated spatial patterns, while also contributing to the other two tasks. 
This demonstrates that task-specific experts focus on their primary task and capture distinct features for different tasks
\textit{(4)} 
Shared experts provide global contextual knowledge across all three tasks, with particularly strong contributions to population prediction, which relies on information from all modalities. 
This indicates that shared experts are crucial for capturing overarching patterns and enabling effective cross-task knowledge transfer.}

\begin{table}[t!]
\renewcommand{\arraystretch}{1.2}
\centering
\setlength{\abovecaptionskip}{1mm}
\caption{Ablation on region and POI embeddings (MAE↓).
}
\label{tab:ablation}
\scalebox{0.98}{
\begin{tabular}{lccc}
\toprule
\textbf{Task} & Carbon & Population & Light intensity\\
\hline
RemoteCLIP(ST) & 0.9406  & 1.0929 	& 0.2390 \\
UrbanCLIP(ST) & 1.0366 &1.0691 &	0.2415 \\
\textit{w/o} Region \& POI & 0.9316	&0.9347	& 0.2354\\
\textit{w/o} POI & 0.8737	&0.9305	&\underline{0.2331}\\
\textit{w/o} Region & \underline{0.8354} &	\underline{0.8850} &	0.2334\\
\textbf{\name} & \textbf{0.7953} & \textbf{0.7775} & \textbf{0.2320}\\
\bottomrule
\end{tabular}}
\vspace{-2mm}
\end{table}



\noindent \textbf{Effect of Embedding}. 
We further examine the effect of region-level and POI embeddings on model performance. To this end, we perform ablation studies by selectively removing these components and evaluating the resulting model variants. Results on the SH dataset are reported in Table~\ref{tab:ablation}. The considered variants include:

\begin{itemize}
\vspace{1mm}
\item {\textbf{UrbanCLIP (ST)}}. Single-task UrbanCLIP baseline.
\item{\textbf{RemoteCLIP (ST)}}. Single-task RemoteCLIP baseline. 
\item{\textbf{w/o Region \& POI}}. We omit both region and POI embeddings.
\item{\textbf{w/o POI}}. We omit the POI embeddings. 
\item{\textbf{w/o Region}}. We omit the region embeddings.
\vspace{1mm}
\end{itemize}

Based on the ablation results, several conclusions can be drawn: 
\textit{(1)} Even the simplest configuration without either embedding surpasses strong baselines (RemoteCLIP and UrbanCLIP) in average performance, demonstrating the effectiveness of our lightweight architecture. 
\textit{(2)} 
Using region or POI embeddings individually enhances performance: 
region embeddings particularly benefit environmental indicators like carbon and light intensity by encoding spatial context, 
while POI embeddings excel in socio-economic tasks like population and carbon prediction through functional semantics.
\textit{(3)} 
Combining both embeddings achieves optimal results with notable gains in population prediction, demonstrating that spatial structure and functional semantics provide complementary information for comprehensive urban representation.

\subsection{Visualization}
In the \textbf{Appendix~\ref{sec:Visualization}}, we provide a detailed analysis of the internal behavior of UrbanMoE through T-SNE visualizations of expert embeddings and an examination of sparse gating weights. 
The T-SNE results reveal clear structural patterns in the learned representation space: shared and dual-task experts tend to form overlapping or adjacent clusters, indicating effective modeling of cross-task and globally transferable spatial–semantic knowledge, whereas task-specific experts exhibit more dispersed distributions, capturing fine-grained and modality-sensitive characteristics unique to individual tasks. 
In addition, the sparse gating analysis highlights strong task-dependent expert preferences. Carbon emission prediction predominantly activates dual-task experts, reflecting strong inter-task correlations; population prediction exhibits a more balanced utilization of shared, dual-task, and task-specific experts; and nighttime light intensity prediction emphasizes modality-specific experts, particularly visual experts for spatial structure and textual experts for semantic cues. 
Across all tasks, non-essential experts are selectively deactivated, demonstrating that UrbanMoE achieves both interpretability and efficiency through adaptive and sparse expert routing.

\section{RELATED WORK}
\subsection{Single-Modal Urban Profiling}
Urban region profiling using a single modality focuses on learning representations of urban areas from one type of data source, such as satellite imagery, POIs or human mobility~\cite{chencross}. Yeh \etal ~\cite{yeh2020using} aggregate satellite image features to predict economic well-being in Africa. Other studies ~\cite{he2018perceiving, park2022learning,jean2019tile2vec} similarly focus solely on extracting information from satellite images, or on modeling relationships between satellite image tiles to capture spatial context. Huang \etal ~\cite{huang2023learning} utilizes the co-occurrence relationships between POIs to construct a graph structure and learns region representations through a hierarchical graph neural network and mutual information maximization. Zhang \etal ~\cite{zhang2021multi} constructs multi-view graphs to more comprehensively capture the complex relationships between regions. Besides, graph-enhanced spatio-temporal large language models, notably ST-LLM+~\cite{11005661}, have emerged as a new trend for effective spatio-temporal prediction via pre-training.

The representational capacity of single-modality approaches is inherently limited by their isolated perspective. 
To create a more holistic characterization of urban systems, we introduce a multimodal framework that amalgamates complementary information from visual, textual, and structural sources. 
\vspace{-2mm}
\subsection{Multimodal Urban Profiling}
Fusing multi-source heterogeneous data can yield a more comprehensive urban profile. In recent years, multi-modal learning has become mainstream. The most typical fusion approach combines satellite and street view imagery. For example, Law \etal ~\cite{law2019take} use both satellite and street view images to estimate house prices, demonstrating the effectiveness of complementary multi-perspective visual information.
Some studies ~\cite{xi2022beyond,huang2021m3g,wang2025multi} focus on combining visual data with structural modalities, where POIs are integrated with satellite or street-view imagery to capture spatial structures and functional semantics.
Other studies ~\cite{yan2024urbanclip,liu2024remoteclip,hao2025urbanvlp,xiao2024refound,chen2024profiling} leverage visual-language fusion, inspired by large-scale vision-language pretraining, to align urban imagery with textual descriptions and enable more interpretable urban representations.
In addition, Chen \etal ~\cite{chen2026region} use POI and human mobility data for region embedding analysis.  

Existing multimodal urban profiling faces limited cross-task generalization and lacks dynamic modality balancing. 
We propose a unified multi-task framework integrating visual, textual, and structural data.
Through complementary fusion and task-aware routing, our method enables adaptive feature selection, achieving robust performance across urban tasks.

\section{CONCLUSION}
Urban region profiling requires integrating heterogeneous data sources and predicting diverse urban indicators.
 However, existing methods fail to effectively model cross-modal and cross-task knowledge transfer, which is essential for practical applications. 
 In this paper, we propose UrbanMoE, a unified framework for adaptive multi-modal and multi-task urban profiling. By leveraging sparse expert selection and disentangled fusion, UrbanMoE enables efficient and interpretable cross-modal knowledge sharing. 
 Experiments on multiple urban datasets show consistent improvements in predicting carbon emissions, population, and light intensity.


While UrbanMoE performs well in the current setting, future work will further extend the framework toward inductive region representations derived from attributes, spatial context, or multimodal observations, enabling stronger generalization to unseen regions and new cities in real-world urban applications.
 

\section{ACKNOWLEDGMENT}
This work is supported by Jilin Province Industrial Key Core Technology Tackling Project (20230201085GX). Zijian Zhang is supported by China Postdoctoral Science Foundation (2025M771587) and the Open Funding Programs of State Key Laboratory of AI Safety. Qingliang Li is supported by the National Natural Science Foundation of China (42575159, 42275155, 62206028). Irwin King is supported by the Research Grants Council of the Hong Kong Special Administrative Region, China (CUHK 2410072, RGC R1015-23) and (CUHK 2300246, RGC C1043-24G).
Jiamiao Liu is supervised by Pingping Liu and Zijian Zhang.

\newpage

\bibliographystyle{ACM-Reference-Format}
\balance
\bibliography{References}

@String{Computing = "Computing" }

@String{Computer = "{IEEE} Computer" }

@article{kramer1991nonlinear,
  title     = {Nonlinear principal component analysis using autoassociative neural networks},
  author    = {Kramer, Mark A},
  journal   = {AIChE journal},
  volume    = {37},
  number    = {2},
  pages     = {233--243},
  year      = {1991},
  publisher = {Wiley Online Library}
}

@inproceedings{chencross,
  title={Cross-City Latent Space Alignment for Consistency Region Embedding},
  author={Chen, Meng and Jia, Hongwei and Li, Zechen and Jia, Wenzhen and Zhao, Kai and Dai, Hongjun and Huang, Weiming},
  booktitle={Forty-second International Conference on Machine Learning}
}

@article{chen2026region,
  title={Region Embedding With Adaptive Correlation Discovery for Predicting Urban Socioeconomic Indicators},
  author={Chen, Meng and Jia, Hongwei and Li, Zechen and Huang, Weiming and Zhao, Kai and Gong, Yongshun and Xu, Haoran and Dai, Hongjun},
  journal={IEEE Transactions on Knowledge and Data Engineering},
  volume={38},
  number={2},
pages={1280--1291},
  year={2026},
  publisher={IEEE}
}

@inproceedings{chen2024profiling,
  title={Profiling urban streets: A semi-supervised prediction model based on street view imagery and spatial topology},
  author={Chen, Meng and Li, Zechen and Huang, Weiming and Gong, Yongshun and Yin, Yilong},
  booktitle={Proceedings of the 30th ACM SIGKDD Conference on Knowledge Discovery and Data Mining},
  pages={319--328},
  year={2024}
}

@article{tipping1999probabilistic,
  title     = {Probabilistic principal component analysis},
  author    = {Tipping, Michael E and Bishop, Christopher M},
  journal   = {Journal of the Royal Statistical Society Series B: Statistical Methodology},
  volume    = {61},
  number    = {3},
  pages     = {611--622},
  year      = {1999},
  publisher = {Oxford University Press}
}

@inproceedings{jean2019tile2vec,
  title     = {Tile2vec: Unsupervised representation learning for spatially distributed data},
  author    = {Jean, Neal and Wang, Sherrie and Samar, Anshul and Azzari, George and Lobell, David and Ermon, Stefano},
  booktitle = {Proceedings of the AAAI Conference on Artificial Intelligence},
  volume    = {33},
  number    = {01},
  pages     = {3967--3974},
  year      = {2019}
}

@inproceedings{he2016deep,
  title     = {Deep residual learning for image recognition},
  author    = {He, Kaiming and Zhang, Xiangyu and Ren, Shaoqing and Sun, Jian},
  booktitle = {Proceedings of the IEEE conference on computer vision and pattern recognition},
  pages     = {770--778},
  year      = {2016}
}

@inproceedings{han2020lightweight,
  title     = {Lightweight and robust representation of economic scales from satellite imagery},
  author    = {Han, Sungwon and Ahn, Donghyun and Cha, Hyunji and Yang, Jeasurk and Park, Sungwon and Cha, Meeyoung},
  booktitle = {Proceedings of the AAAI Conference on Artificial Intelligence},
  volume    = {34},
  number    = {01},
  pages     = {428--436},
  year      = {2020}
}

@inproceedings{xi2022beyond,
  title     = {Beyond the first law of geography: Learning representations of satellite imagery by leveraging point-of-interests},
  author    = {Xi, Yanxin and Li, Tong and Wang, Huandong and Li, Yong and Tarkoma, Sasu and Hui, Pan},
  booktitle = {Proceedings of the ACM web conference 2022},
  pages     = {3308--3316},
  year      = {2022}
}

@article{liu2024remoteclip,
  title     = {Remoteclip: A vision language foundation model for remote sensing},
  author    = {Liu, Fan and Chen, Delong and Guan, Zhangqingyun and Zhou, Xiaocong and Zhu, Jiale and Ye, Qiaolin and Fu, Liyong and Zhou, Jun},
  journal   = {IEEE Transactions on Geoscience and Remote Sensing},
  volume    = {62},
  pages     = {1--16},
  year      = {2024},
  publisher = {IEEE}
}

@inproceedings{yan2024urbanclip,
  title     = {Urbanclip: Learning text-enhanced urban region profiling with contrastive language-image pretraining from the web},
  author    = {Yan, Yibo and Wen, Haomin and Zhong, Siru and Chen, Wei and Chen, Haodong and Wen, Qingsong and Zimmermann, Roger and Liang, Yuxuan},
  booktitle = {Proceedings of the ACM Web Conference 2024},
  pages     = {4006--4017},
  year      = {2024}
}

@inproceedings{chen2020simple,
  title        = {A simple framework for contrastive learning of visual representations},
  author       = {Chen, Ting and Kornblith, Simon and Norouzi, Mohammad and Hinton, Geoffrey},
  booktitle    = {International conference on machine learning},
  pages        = {1597--1607},
  year         = {2020},
  organization = {PmLR}
}

@article{tatem2017worldpop,
  title     = {WorldPop, open data for spatial demography},
  author    = {Tatem, Andrew J},
  journal   = {Scientific data},
  volume    = {4},
  number    = {1},
  pages     = {1--4},
  year      = {2017},
  publisher = {Nature Publishing Group}
}

@inproceedings{he2018perceiving,
  title     = {Perceiving commerial activeness over satellite images},
  author    = {He, Zhiyuan and Yang, Su and Zhang, Weishan and Zhang, Jiulong},
  booktitle = {Companion Proceedings of the The Web Conference 2018},
  pages     = {387--394},
  year      = {2018}
}

@article{oda2018open,
  title     = {The Open-source Data Inventory for Anthropogenic CO 2, version 2016 (ODIAC2016): a global monthly fossil fuel CO 2 gridded emissions data product for tracer transport simulations and surface flux inversions},
  author    = {Oda, Tomohiro and Maksyutov, Shamil and Andres, Robert J},
  journal   = {Earth System Science Data},
  volume    = {10},
  number    = {1},
  pages     = {87--107},
  year      = {2018},
  publisher = {Copernicus GmbH}
}

@article{ecosystem2023copernicus,
  title   = {Copernicus data space ecosystem},
  author  = {Ecosystem, Copernicus Data Space},
  journal = {URL: https://dataspace.copernicus.eu},
  year    = {2023}
}

@article{team2023gemini,
  title   = {Gemini: a family of highly capable multimodal models},
  author  = {Team, Gemini and Anil, Rohan and Borgeaud, Sebastian and Alayrac, Jean-Baptiste and Yu, Jiahui and Soricut, Radu and Schalkwyk, Johan and Dai, Andrew M and Hauth, Anja and Millican, Katie and others},
  journal = {arXiv preprint arXiv:2312.11805},
  year    = {2023}
}

@article{wang2024urbandatalayer,
  title   = {UrbanDataLayer: A Unified Data Pipeline for Urban Science},
  author  = {Wang, Yiheng and Wang, Tianyu and Zhang, YuYing and Zhang, Hongji and Zheng, Haoyu and Zheng, Guanjie and Kong, Linghe},
  journal = {Advances in Neural Information Processing Systems},
  volume  = {37},
  pages   = {7296--7310},
  year    = {2024}
}

@article{elvidge2021annual,
  title     = {Annual time series of global VIIRS nighttime lights derived from monthly averages: 2012 to 2019},
  author    = {Elvidge, Christopher D and Zhizhin, Mikhail and Ghosh, Tilottama and Hsu, Feng-Chi and Taneja, Jay},
  journal   = {Remote Sensing},
  volume    = {13},
  number    = {5},
  pages     = {922},
  year      = {2021},
  publisher = {MDPI}
}

@article{yeh2020using,
  title     = {Using publicly available satellite imagery and deep learning to understand economic well-being in Africa},
  author    = {Yeh, Christopher and Perez, Anthony and Driscoll, Anne and Azzari, George and Tang, Zhongyi and Lobell, David and Ermon, Stefano and Burke, Marshall},
  journal   = {Nature communications},
  volume    = {11},
  number    = {1},
  pages     = {2583},
  year      = {2020},
  publisher = {Nature Publishing Group UK London}
}

@inproceedings{park2022learning,
  title     = {Learning economic indicators by aggregating multi-level geospatial information},
  author    = {Park, Sungwon and Han, Sungwon and Ahn, Donghyun and Kim, Jaeyeon and Yang, Jeasurk and Lee, Susang and Hong, Seunghoon and Kim, Jihee and Park, Sangyoon and Yang, Hyunjoo and others},
  booktitle = {Proceedings of the AAAI Conference on Artificial Intelligence},
  volume    = {36},
  number    = {11},
  pages     = {12053--12061},
  year      = {2022}
}

@article{huang2023learning,
  title     = {Learning urban region representations with POIs and hierarchical graph infomax},
  author    = {Huang, Weiming and Zhang, Daokun and Mai, Gengchen and Guo, Xu and Cui, Lizhen},
  journal   = {ISPRS Journal of Photogrammetry and Remote Sensing},
  volume    = {196},
  pages     = {134--145},
  year      = {2023},
  publisher = {Elsevier}
}

@inproceedings{zhang2021multi,
  title     = {Multi-view joint graph representation learning for urban region embedding},
  author    = {Zhang, Mingyang and Li, Tong and Li, Yong and Hui, Pan},
  booktitle = {Proceedings of the twenty-ninth international conference on international joint conferences on artificial intelligence},
  pages     = {4431--4437},
  year      = {2021}
}

@article{law2019take,
  title     = {Take a look around: using street view and satellite images to estimate house prices},
  author    = {Law, Stephen and Paige, Brooks and Russell, Chris},
  journal   = {ACM Transactions on Intelligent Systems and Technology (TIST)},
  volume    = {10},
  number    = {5},
  pages     = {1--19},
  year      = {2019},
  publisher = {ACM New York, NY, USA}
}

@article{huang2021m3g,
  title   = {M3G: Learning urban neighborhood representation from multi-modal multi-graph},
  author  = {Huang, Tianyuan and Wang, Zhecheng and Sheng, Hao and Ng, Andrew Y and Rajagopal, Ram},
  journal = {Proceedings of the DeepSpatial},
  volume  = {2021},
  pages   = {2nd},
  year    = {2021}
}

@inproceedings{hao2025urbanvlp,
  title     = {Urbanvlp: Multi-granularity vision-language pretraining for urban socioeconomic indicator prediction},
  author    = {Hao, Xixuan and Chen, Wei and Yan, Yibo and Zhong, Siru and Wang, Kun and Wen, Qingsong and Liang, Yuxuan},
  booktitle = {Proceedings of the AAAI Conference on Artificial Intelligence},
  volume    = {39},
  number    = {27},
  pages     = {28061--28069},
  year      = {2025}
}

@inproceedings{xiao2024refound,
  title     = {Refound: Crafting a foundation model for urban region understanding upon language and visual foundations},
  author    = {Xiao, Congxi and Zhou, Jingbo and Xiao, Yixiong and Huang, Jizhou and Xiong, Hui},
  booktitle = {Proceedings of the 30th ACM SIGKDD Conference on Knowledge Discovery and Data Mining},
  pages     = {3527--3538},
  year      = {2024}
}

@inproceedings{liu2023knowledge,
  title     = {Knowledge-infused contrastive learning for urban imagery-based socioeconomic prediction},
  author    = {Liu, Yu and Zhang, Xin and Ding, Jingtao and Xi, Yanxin and Li, Yong},
  booktitle = {Proceedings of the ACM web conference 2023},
  pages     = {4150--4160},
  year      = {2023}
}

@article{wang2025multi,
  title     = {Multi-modal contrastive learning of urban space representations from POI data},
  author    = {Wang, Xinglei and Cheng, Tao and Law, Stephen and Zeng, Zichao and Yin, Lu and Liu, Junyuan},
  journal   = {Computers, Environment and Urban Systems},
  volume    = {120},
  pages     = {102299},
  year      = {2025},
  publisher = {Elsevier}
}

@article{abitbol2020interpretable,
  title     = {Interpretable socioeconomic status inference from aerial imagery through urban patterns},
  author    = {Abitbol, Jacob Levy and Karsai, Marton},
  journal   = {Nature Machine Intelligence},
  volume    = {2},
  number    = {11},
  pages     = {684--692},
  year      = {2020},
  publisher = {Nature Publishing Group UK London}
}

@article{gebru2017using,
  title     = {Using deep learning and Google Street View to estimate the demographic makeup of neighborhoods across the United States},
  author    = {Gebru, Timnit and Krause, Jonathan and Wang, Yilun and Chen, Duyun and Deng, Jia and Aiden, Erez Lieberman and Fei-Fei, Li},
  journal   = {Proceedings of the National Academy of Sciences},
  volume    = {114},
  number    = {50},
  pages     = {13108--13113},
  year      = {2017},
  publisher = {National Academy of Sciences}
}

@article{jean2016combining,
  title     = {Combining satellite imagery and machine learning to predict poverty},
  author    = {Jean, Neal and Burke, Marshall and Xie, Michael and Alampay Davis, W Matthew and Lobell, David B and Ermon, Stefano},
  journal   = {Science},
  volume    = {353},
  number    = {6301},
  pages     = {790--794},
  year      = {2016},
  publisher = {American Association for the Advancement of Science}
}

@article{he2020population,
  title     = {Population spatialization in Beijing city based on machine learning and multisource remote sensing data},
  author    = {He, Miao and Xu, Yongming and Li, Ning},
  journal   = {Remote Sensing},
  volume    = {12},
  number    = {12},
  pages     = {1910},
  year      = {2020},
  publisher = {MDPI}
}

@article{wang2022population,
  title     = {A population spatialization model at the building scale using random forest},
  author    = {Wang, Mengqi and Wang, Yinglin and Li, Bozhao and Cai, Zhongliang and Kang, Mengjun},
  journal   = {remote sensing},
  volume    = {14},
  number    = {8},
  pages     = {1811},
  year      = {2022},
  publisher = {MDPI}
}

@article{lu2017predicting,
  title     = {Predicting transportation carbon emission with urban big data},
  author    = {Lu, Xiangyong and Ota, Kaoru and Dong, Mianxiong and Yu, Chen and Jin, Hai},
  journal   = {IEEE Transactions on Sustainable Computing},
  volume    = {2},
  number    = {4},
  pages     = {333--344},
  year      = {2017},
  publisher = {IEEE}
}

@article{naz2023comparative,
  title     = {Comparative analysis of deep learning and statistical models for air pollutants prediction in urban areas},
  author    = {Naz, Fareena and Mccann, Conor and Fahim, Muhammad and Cao, Tuan-Vu and Hunter, Ruth and Viet, Nguyen Trung and Nguyen, Long D and Duong, Trung Q},
  journal   = {IEEE Access},
  volume    = {11},
  pages     = {64016--64025},
  year      = {2023},
  publisher = {IEEE}
}

@inproceedings{hao2025nature,
  title     = {Nature makes no leaps: Building continuous location embeddings with satellite imagery from the web},
  author    = {Hao, Xixuan and Chen, Wei and Zou, Xingchen and Liang, Yuxuan},
  booktitle = {Proceedings of the ACM on Web Conference 2025},
  pages     = {2799--2812},
  year      = {2025}
}

@inproceedings{ma2018modeling,
  title     = {Modeling task relationships in multi-task learning with multi-gate mixture-of-experts},
  author    = {Ma, Jiaqi and Zhao, Zhe and Yi, Xinyang and Chen, Jilin and Hong, Lichan and Chi, Ed H},
  booktitle = {Proceedings of the 24th ACM SIGKDD international conference on knowledge discovery \& data mining},
  pages     = {1930--1939},
  year      = {2018}
}

@article{jacobs1991adaptive,
  title     = {Adaptive mixtures of local experts},
  author    = {Jacobs, Robert A and Jordan, Michael I and Nowlan, Steven J and Hinton, Geoffrey E},
  journal   = {Neural computation},
  volume    = {3},
  number    = {1},
  pages     = {79--87},
  year      = {1991},
  publisher = {MIT Press}
}

@ARTICLE{11005661,
  author={Liu, Chenxi and Hettige, Kethmi Hirushini and Xu, Qianxiong and Long, Cheng and Xiang, Shili and Cong, Gao and Li, Ziyue and Zhao, Rui},
  journal={TKDE}, 
  title={ST-LLM+: Graph Enhanced Spatio-Temporal Large Language Models for Traffic Prediction}, 
  year={2025},
  volume={37},
  number={8},
  pages={4846-4859}}

\appendix

\section{Baseline Methods}
\label{baseline}

\begin{itemize}
\item {\textbf{Autoencoder}}~\cite{kramer1991nonlinear}. A neural network architecture that learns feature representations from unlabeled satellite images by minimizing reconstruction error. It serves as a nonlinear dimensionality reduction method for urban prediction tasks. 
\item{\textbf{PCA}}~\cite{tipping1999probabilistic}. Principal Component Analysis is applied to transform raw satellite imagery into extended feature vectors. The top principal components are selected to retain the majority of variance in the data. 
\item{\textbf{ResNet-18}}~\cite{he2016deep}. A widely used convolutional neural network pre-trained on ImageNet, directly transferred to extract satellite image features for socioeconomic indicator prediction. 
\item{\textbf{Tile2Vec}}~\cite{jean2019tile2vec}. An unsupervised geospatial representation learning method based on a triplet loss, which minimizes the distance between embeddings of neighboring satellite tiles while maximizing the distance from distant ones. 
\item{\textbf{READ}}~\cite{han2020lightweight}. Representation Extraction over an Arbitrary District is a semi-supervised framework that leverages limited labeled samples in combination with transfer learning. It adopts a knowledge distillation strategy, where a pre-trained teacher network guides a lightweight student model, enabling the extraction of compact yet robust satellite image representations.
\item{\textbf{PG-SimCLR}}~\cite{xi2022beyond}. A contrastive learning approach for geospatial imagery that extends SimCLR~\cite{chen2020simple}. It enforces similar representations for spatially adjacent or functionally related regions, thereby improving the quality of unsupervised urban feature learning.
\item{\textbf{UrbanCLIP}}~\cite{yan2024urbanclip}. A multimodal pretraining framework designed for urban region profiling using cross-modal supervision from web-scale data. Performance is reported under both single-task and multi-task settings to evaluate its generalization ability across urban socioeconomic indicators. 
\item{\textbf{RemoteCLIP}}~\cite{liu2024remoteclip}. A multimodal pretraining model trained on large-scale remote sensing image–text pairs. Similar to UrbanCLIP, both single-task and multi-task results are reported to benchmark its capacity for capturing multimodal urban semantics.
\end{itemize}

\section{Expert Allocation Strategy}
\label{Expert Allocation}

\begin{table}[htbp]
\centering
\caption{Expert allocation for different configurations (total 16 experts per configuration).}
\label{tab:expert_allocation}
\begin{tabular}{lccc}
\toprule
Configuration & Special-task & Shared & Dual-task (per task) \\
\midrule
Special-task only & 16 & 0 & 0 \\
Shared only & 0 & 16 & 0 \\
Dual-task only & 0 & 0 & 8--8 \\
Special + Shared & 8 & 8 & 0 \\
Special + Dual & 8 & 0 & 4--4 \\
Shared + Dual & 0 & 8 & 4--4 \\
All three types & 8 & 4 & 2--2 \\
\bottomrule
\end{tabular}
\end{table}

Table~\ref{tab:expert_allocation} summarizes the allocation of experts for each ablation configuration used in Section~\ref{sec:Effect of Expert Types}. 
For fair comparison, the total number of experts is fixed to 16 in every configuration. 
The allocation is distributed among three types of experts: special-task, shared, and dual-task (with dual-task experts evenly split between the two related tasks). 
This setup allows us to systematically evaluate the contribution of each expert type and their combinations to the overall prediction performance.

This detailed allocation enables readers to reproduce the ablation studies and understand how the distribution of experts influences the per-task and overall model performance.

\section{Hyper-Parameter Analysis}

\noindent In this subsection, we analyze the impact of key hyper-parameters on the performance of \name using the SH dataset.
First, we vary the dimension of region embeddings $\boldsymbol{r}$ from \{2, 4, 6, 8, 10\}, and present the results in Figure ~\ref{fig:Hyper} (a).
As the embedding dimension increases from 2 to 6, performance steadily improves across all three tasks, with the best results achieved at 6 dimensions.  However, further increasing the dimension beyond 6 leads to slight performance degradation, suggesting that excessively high-dimensional embeddings may introduce redundancy and overfitting.
Next, we investigate the effect of the gating weight threshold $\epsilon $, varying it from \{0, 0.01, 0.02, 0.03, 0.04\}, as shown in Figure ~\ref{fig:Hyper} (b). The results indicate that a small threshold of 0.01 achieves the best performance, whereas larger thresholds tend to suppress expert contributions excessively, resulting in lower task performance. A threshold of 0 still yields competitive results, demonstrating the model’s flexibility in adjusting expert activations without strict thresholding.
Overall, these hyper-parameter studies on the SH dataset show that a moderate region embedding dimension and a small gating weight threshold together provide a good trade-off between representation capacity and generalization.

\label{sec:Hyper-Parameter}
\begin{figure}[htbp]
    \centering
    \includegraphics[width=1\linewidth]{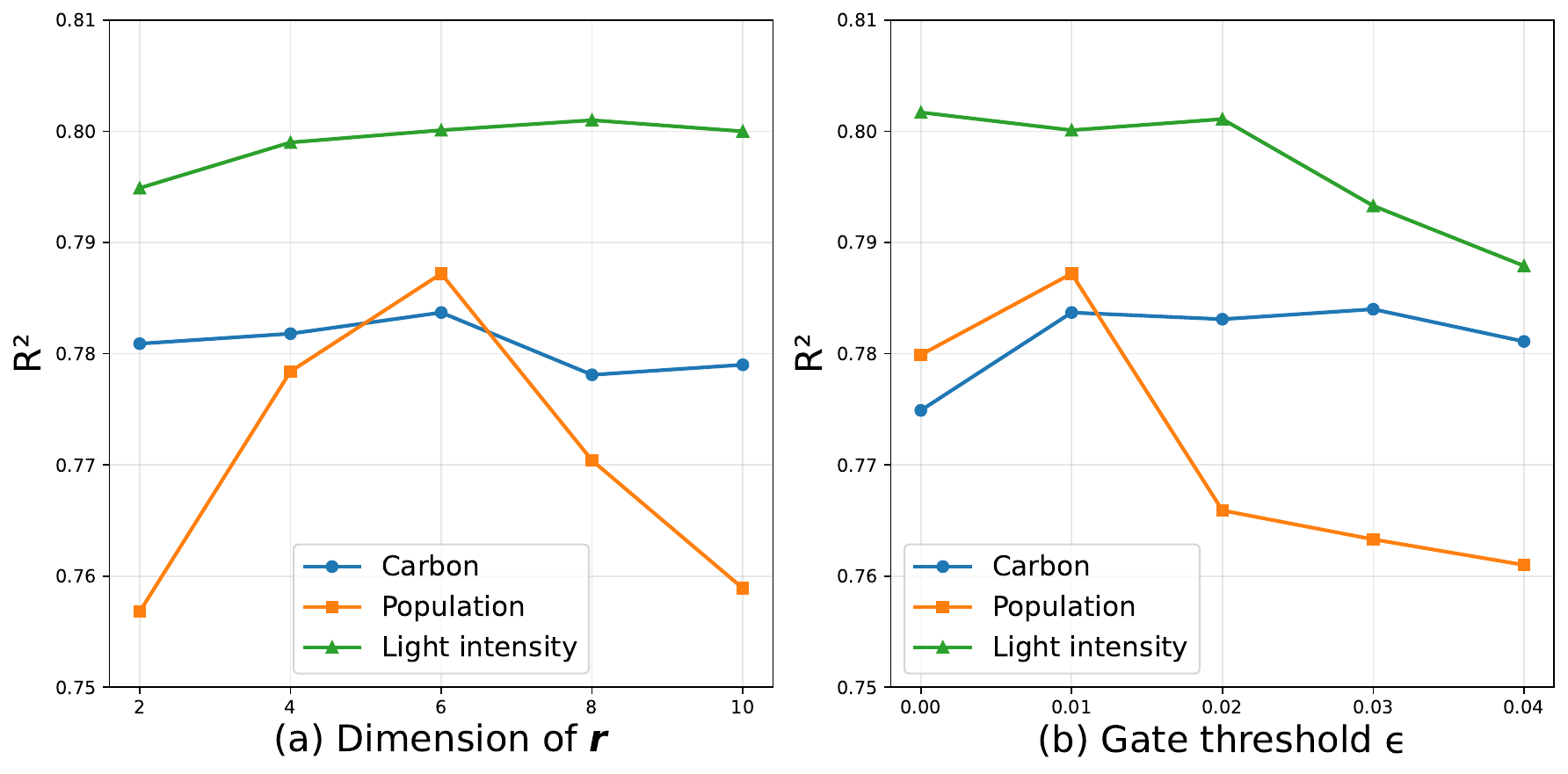}
    \caption{Hyper-parameter analysis results.}
    \label{fig:Hyper}
\end{figure}

\section{Visualization}
\label{sec:Visualization}
\subsection{Visualization of T-SNE}
To further analyze the internal representation learned by UrbanMoE, we visualize the expert outputs using t-SNE, as shown in Figure ~\ref{fig:vis}. 
The subfigures (a) and (b) correspond to the carbon emission task for the image and text branches, respectively;
(c) and (d) correspond to the population task;
and (e) and (f) correspond to the nighttime light intensity task.
Each subfigure illustrates the 2D embedding distribution of samples for one task, separated by modality (image or text). 
The colors correspond to different expert categories, including shared experts, dual-task experts, task-specific experts, and final weighted outputs.

Based on the T-SNE visualizations, several conclusions can be drawn:
\textit{(1)} Shared and dual-task embeddings often overlap or form adjacent clusters, reflecting the strong synergy between global knowledge and inter-task correlations. 
This is expected since both types of experts contribute to modeling shared spatial–semantic patterns that are beneficial to multiple urban indicators.  
\textit{(2)} The specific experts do not always match well with shared or dual-task experts, particularly in tasks such as carbon emissions and nighttime light intensity.
This divergence arises because specific and shared/dual-task experts learn from different perspectives of the input data.
The specific experts focus on fine-grained, task-dependent, and modality-sensitive representations, while the shared and dual-task experts capture more global and transferable spatial-semantic knowledge.
As a result, locally dominant features such as land-use density, or POI-related semantics cannot be fully represented by shared or cross-task experts, leading to partially separated clusters in the embedding space.
\textit{(3)} The final fused outputs integrate knowledge from all three types of experts, but their distribution varies across tasks, indicating different fusion preferences. For example, in the carbon emission prediction task, the final outputs are distributed closer to the shared and dual-task clusters, implying a stronger reliance on cross-task and globally consistent knowledge. 
This observation aligns well with the findings from \textbf{Section ~\ref{sec:Visualization expert weights}}, where carbon emissions were shown to assign higher weights to dual-task experts, further validating the consistency between expert routing behavior and the learned embedding space.

\subsection{Visualization of expert weights}
\label{sec:Visualization expert weights}
To further examine the sparse gating mechanism within UrbanMoE, we visualize the task-specific expert weights on the SH dataset, as shown in Figure~\ref{fig:weight}. Subfigures (a)–(c) illustrate the expert weight distributions for the image branch, while (d)–(f) show those for the text branch, corresponding respectively to carbon emissions, population, and nighttime light intensity.

{Based on the expert weight visualizations, several conclusions can be drawn:
\textit{(1)} 
For carbon emission prediction, the Carbon-Population and Carbon-Light dual-task experts receive the highest weights, reflecting strong cross-task dependencies.
Sparse selection is apparent, with one shared and one specific expert in the image branch deactivated, and one Carbon-Population expert in the text branch zeroed. 
This demonstrates that the model focuses on the most relevant experts while pruning less informative ones.
\textit{(2)} The population prediction task exhibits a more balanced weight distribution across the three types of experts, with no experts deactivated.
This suggests that population modeling benefits from global semantic knowledge captured by shared experts, fine-grained information from task-specific experts, and cross-task correlations via dual-task experts.
\textit{(3)} 
For the nighttime light intensity task, a clear differentiation between image and text branches is observed. 
In the image branch, many experts are deactivated, notably all Carbon-Light dual-task experts, while shared and task-specific experts dominate, reflecting the importance of global spatial structures and localized visual features. 
In the text branch, task-specific experts carry more weight, highlighting the importance of POI-derived semantic information. 
\textit{(4)} 
Across tasks, expert allocation matches Section~\ref{sec:Effect of Expert Types}:  carbon relies on dual-task experts, population distributes attention evenly, and nighttime light emphasizes modality-specific features. 
Sparse selection further improves efficiency by activating only the most relevant experts, while preserving cross-task and modality-specific knowledge for accurate and interpretable predictions.
}

\begin{figure}[!t]
{\subfigure{\includegraphics[width=0.49\linewidth]{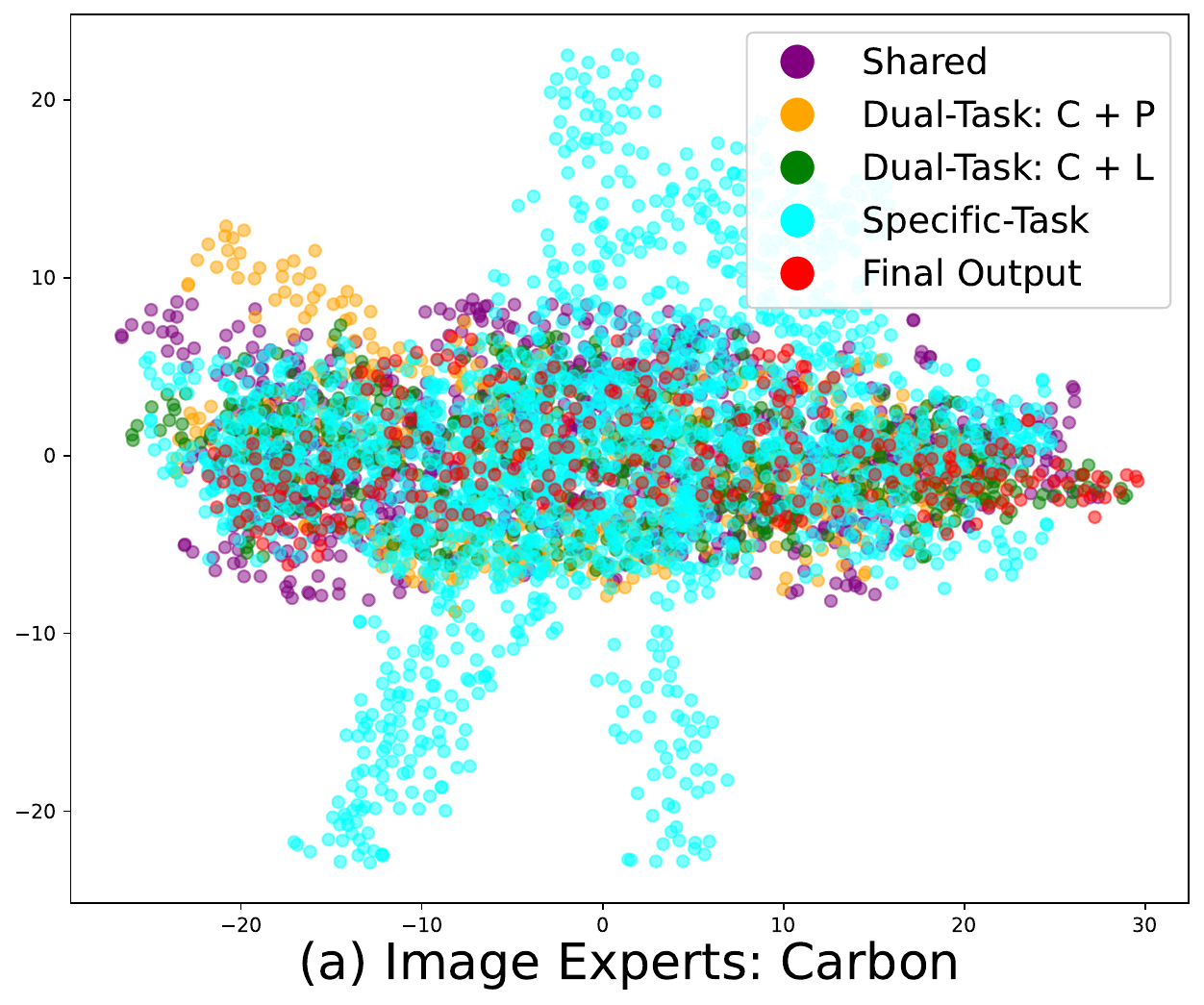}}}
{\subfigure{\includegraphics[width=0.49\linewidth]{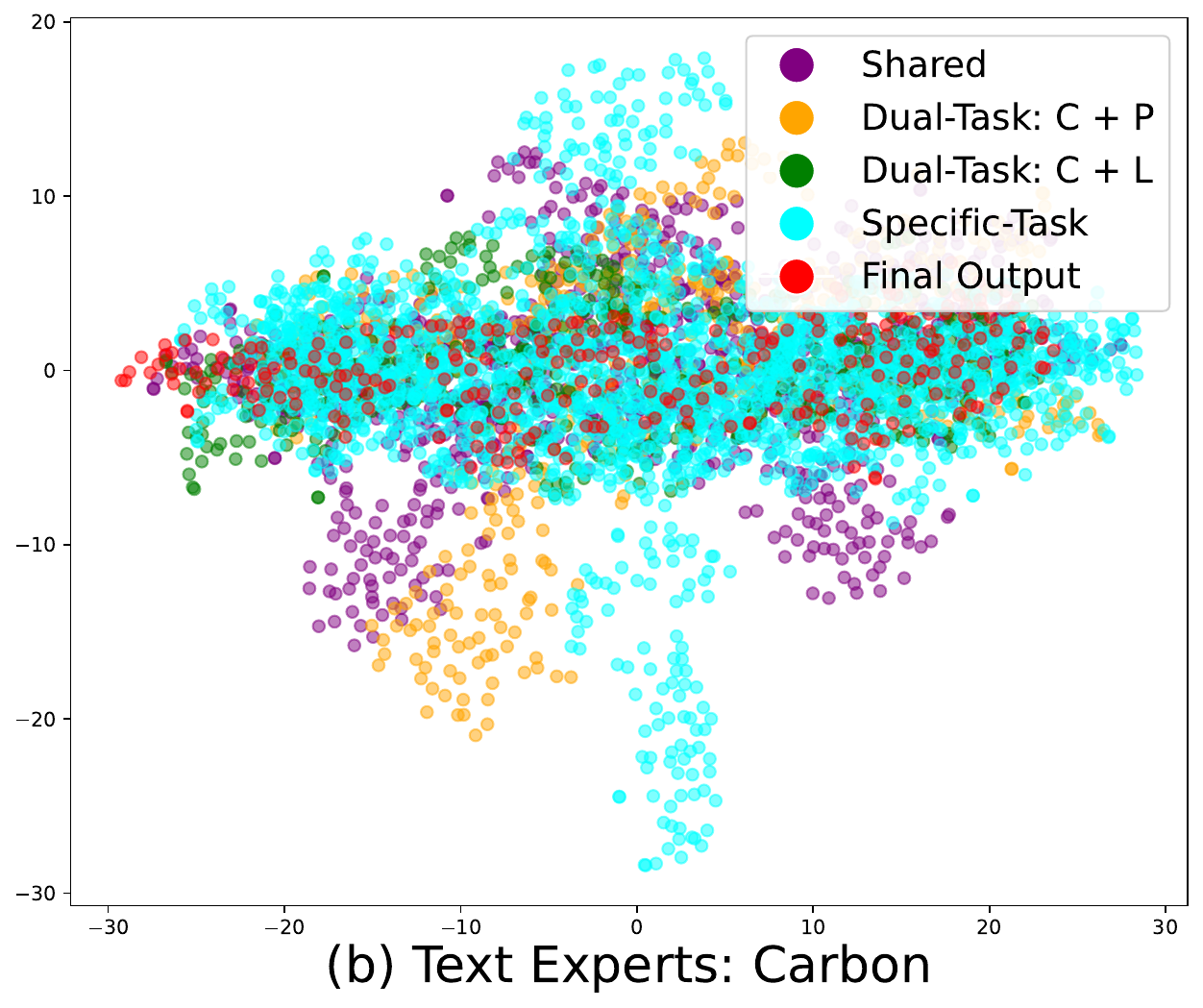}}}\\
{\subfigure{\includegraphics[width=0.49\linewidth]{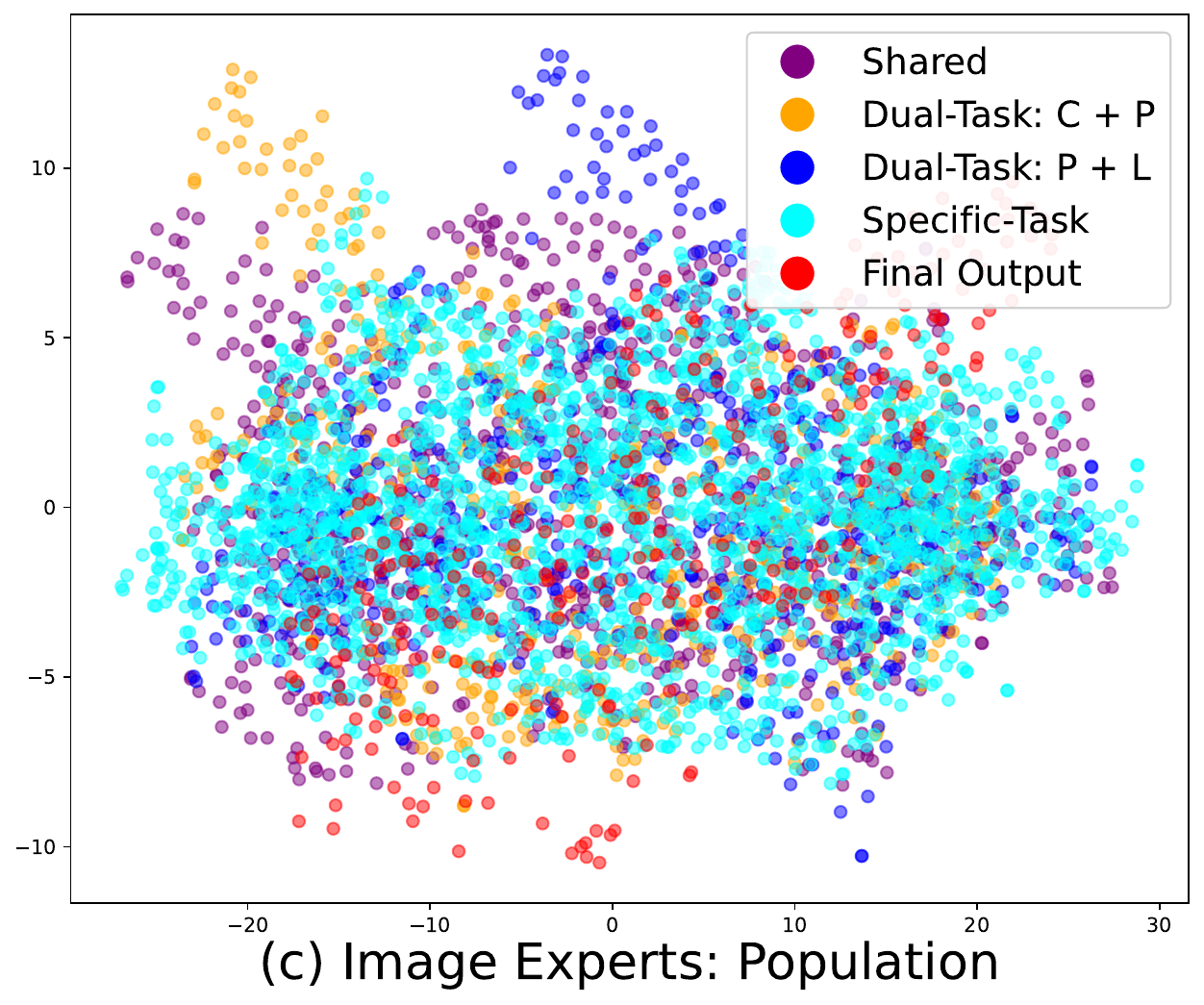}}}
{\subfigure{\includegraphics[width=0.49\linewidth]{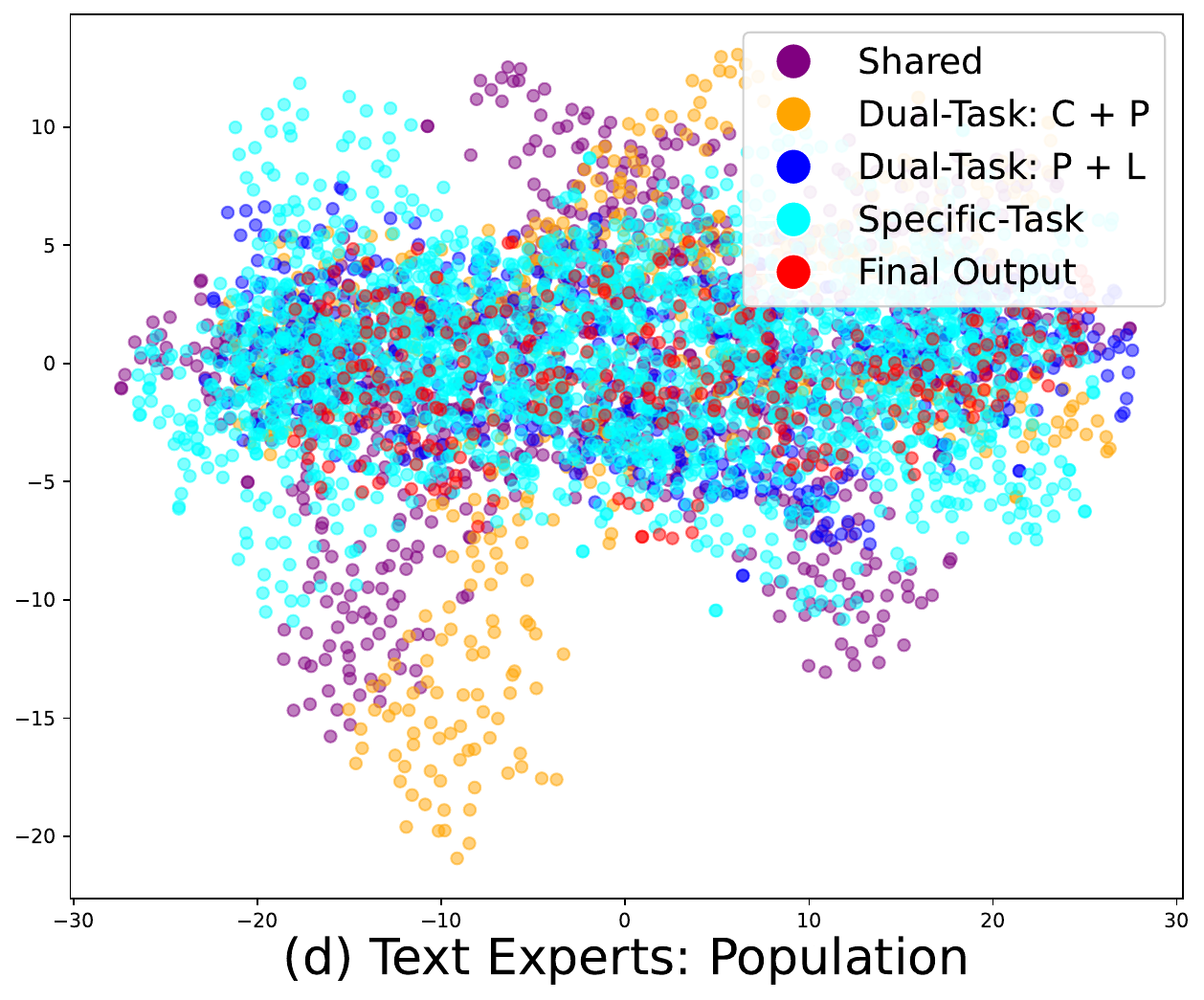}}}\\
{\subfigure{\includegraphics[width=0.49\linewidth]{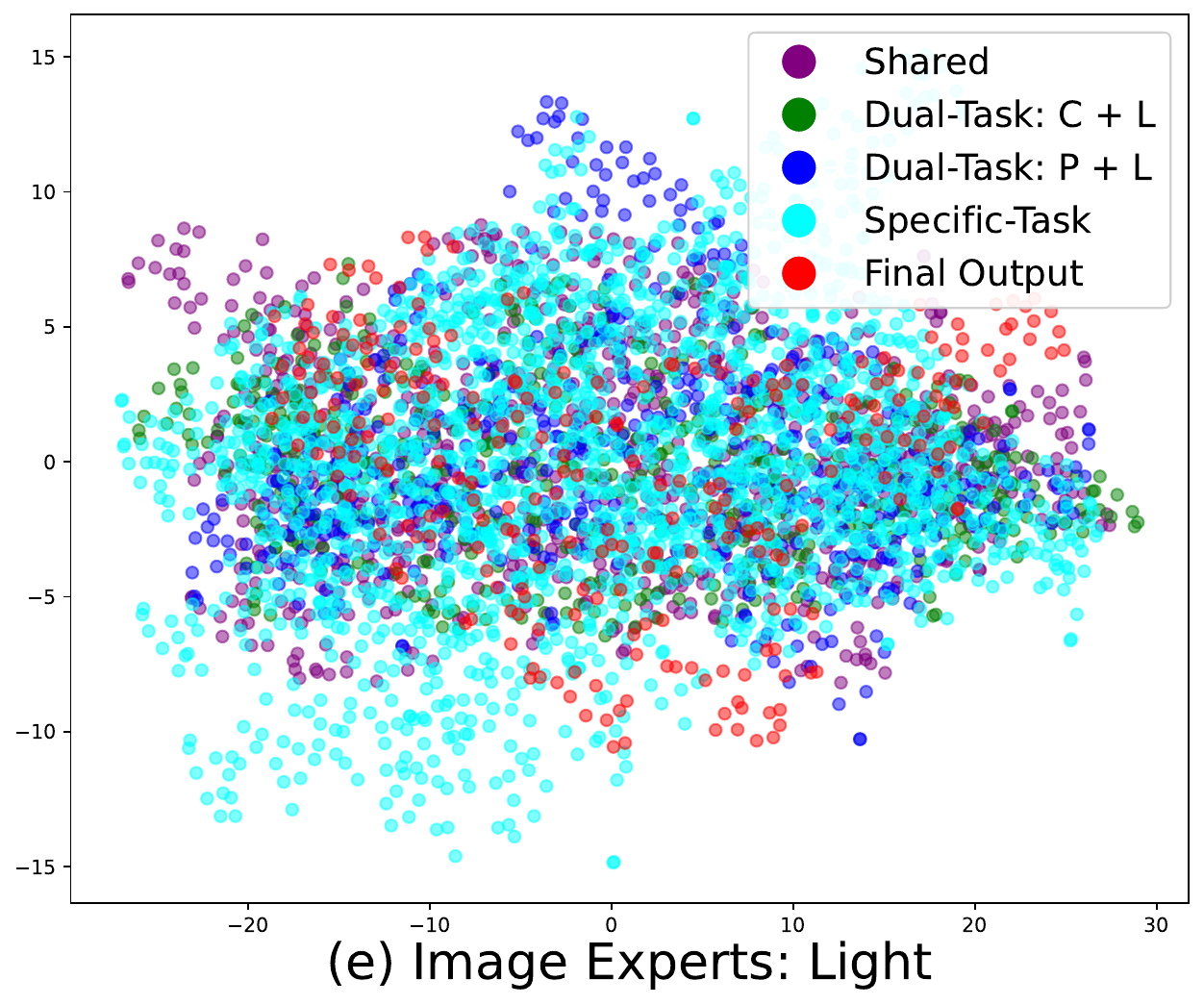}}}
{\subfigure{\includegraphics[width=0.49\linewidth]{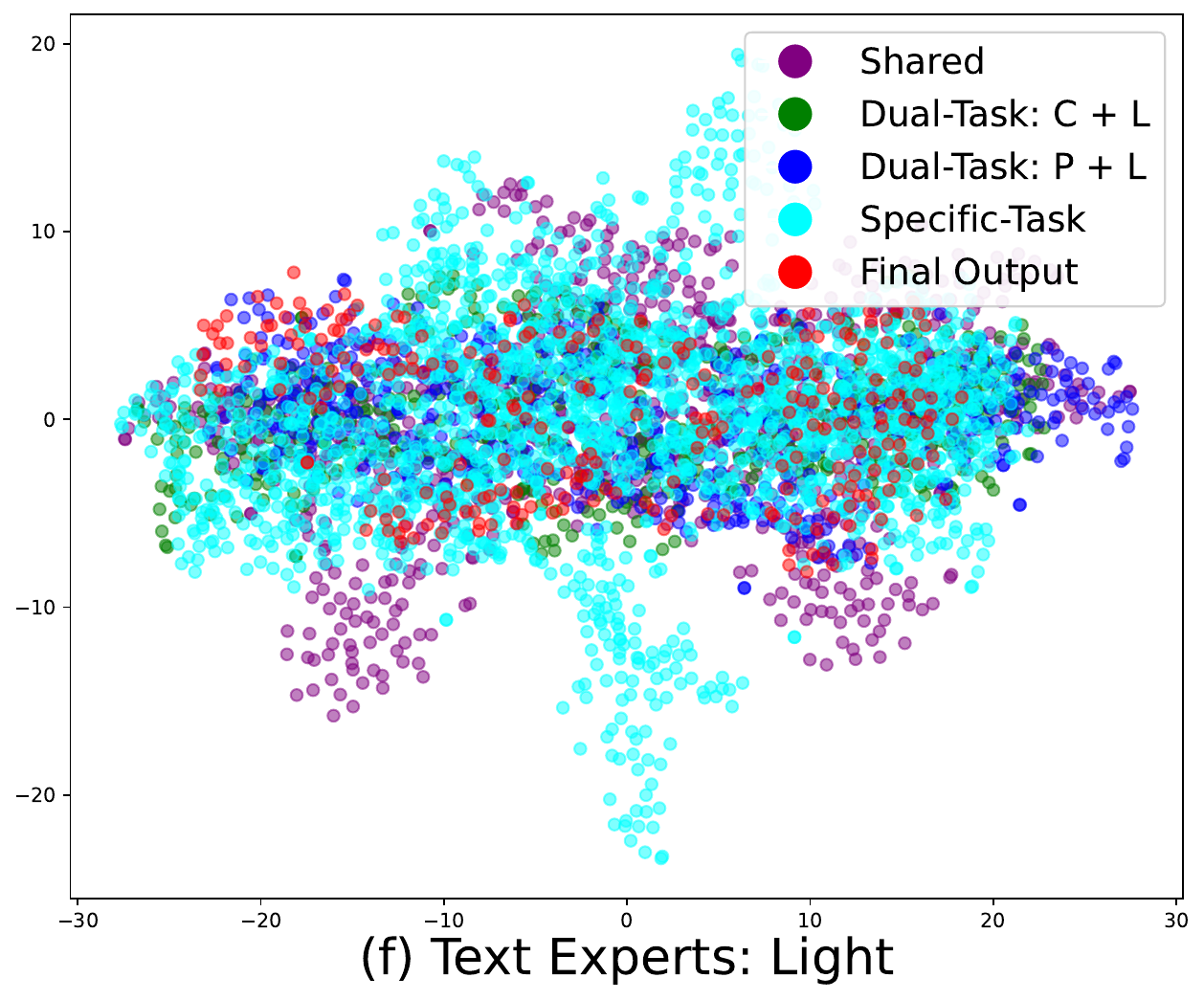}}}
    \caption{T-SNE visualization of multimodal expert embeddings across tasks and modalities.}
    \label{fig:vis}
    \vspace{-3mm}
\end{figure}

\begin{figure*}[!t]
    \setlength{\abovecaptionskip}{-1mm}
    \setlength{\belowcaptionskip}{-3mm}
    \centering
    \includegraphics[width=1\linewidth]{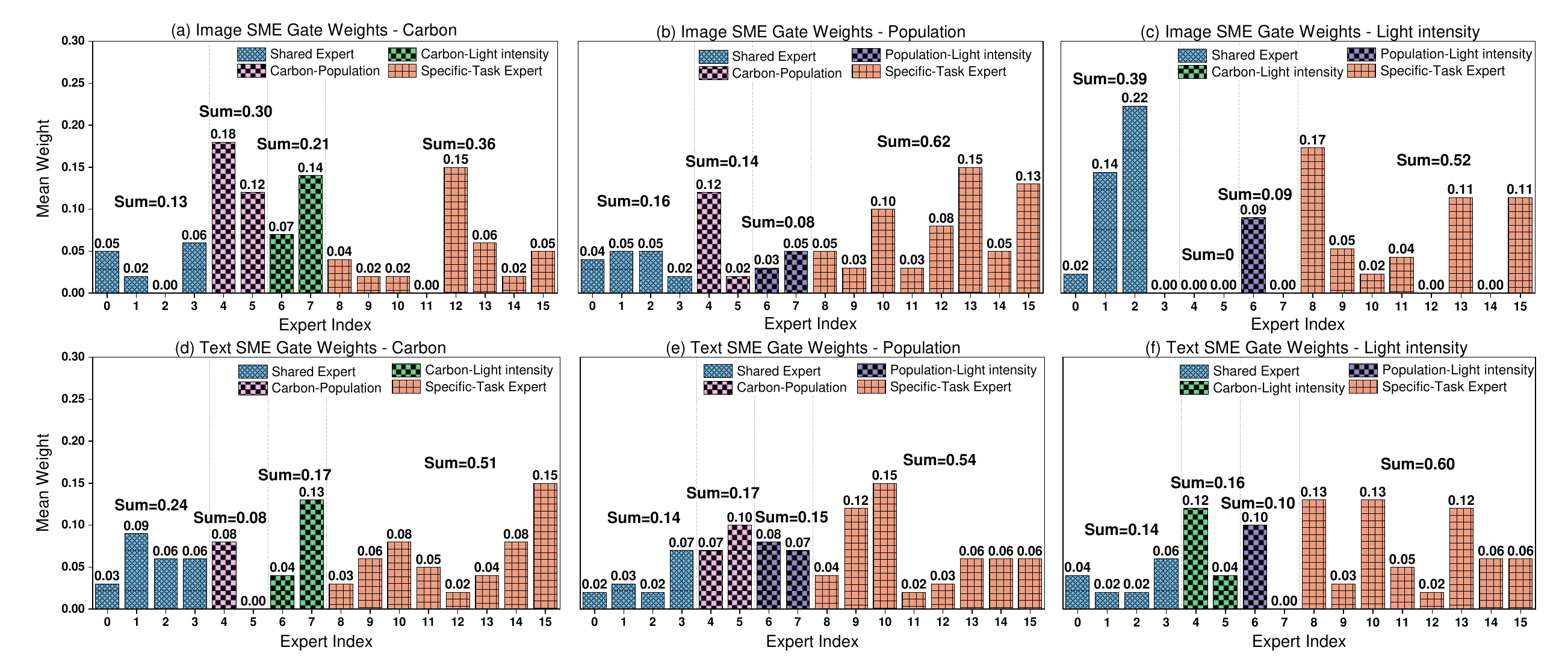}
    \caption{Visualization of sparse gating routing weights.}
    \label{fig:weight}
\end{figure*}

\begin{figure*}[!t]
{{\includegraphics[width=1\linewidth]{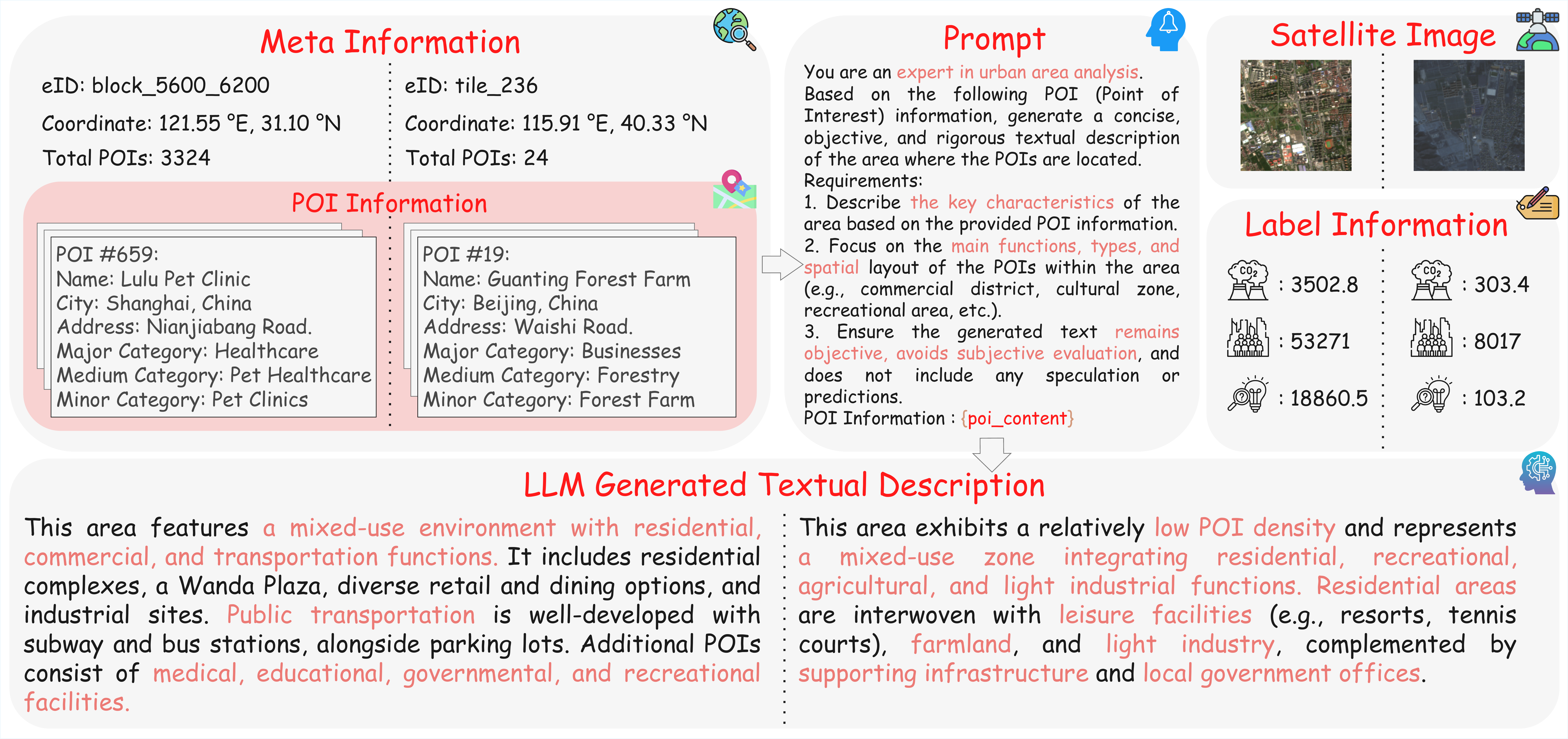}}}
\caption{Example of data sample and LLM prompt generation process.}
    \label{datamark}
\end{figure*}

\section{Data sample and LLM prompt}
\label{sec:LLM prompt}
To construct a multimodal urban profiling dataset, we integrate heterogeneous data sources, including remote-sensing imagery, POI information, and urban indicator labels. 
Figure ~\ref{datamark} presents an example of a data sample, the corresponding LLM prompt, and the generated textual description.
The figure illustrates the process of generating textual descriptions from the original structured data,
specifically meta information and POI lists, via an LLM-based prompt.

\vspace{1mm}
\noindent\textbf{Data sources. }
Each urban region is represented by three primary sources:

\begin{itemize}
    \item \textbf{Satellite imagery}, captured from Sentinel-2~\cite{ecosystem2023copernicus} optical data and street-level imagery, providing spatial and morphological information such as land cover, built-up density, and illumination patterns.
    \item \textbf{POI data}, collected from ~\cite{wang2024urbandatalayer}, which include detailed attributes such as POI name, category, address, and function type. These POIs reflect the socio-economic composition and functional diversity of each region.
    \item \textbf{Label information}, consisting of three quantitative indicators representing key aspects of urban sustainability and development: \textit{Carbon emissions}~\cite{oda2018open} (tonnes of CO$_2$ per region), \textit{Population}~\cite{tatem2017worldpop} (number of Population per region), \textit{Nighttime light intensity}~\cite{elvidge2021annual} (radiance value per region).
\end{itemize}

\vspace{1mm}
\noindent\textbf{Multimodal data construction. } 
For each region, the above data sources are aligned to build three complementary modalities:
\textit{the image modality} derived from satellite imagery, \textit{the POI modality} represented by structured categorical and spatial features, and \textit{the text modality} generated automatically by a LLM based on a carefully designed prompt that converts structured POI information into natural language.


\vspace{2mm}
\noindent Overall, the constructed dataset forms a tri-modal representation that combines image, POI, and text information with three regression targets, including carbon emissions, population, and nighttime light intensity. This dataset serves as the foundation for multimodal and multi-task urban analysis.

\end{document}